\documentclass[11pt,notitlepage,a4paper]{article}

\usepackage[margin=1in]{geometry}
\usepackage{mwe}
\usepackage{graphicx}
\usepackage{epsfig}
\usepackage{color}
\usepackage{hyperref}
\usepackage{amssymb,amsmath}
\numberwithin{equation}{section}
\usepackage{adjustbox}
\usepackage{multirow}
\usepackage{makecell}
\usepackage{cite}
\usepackage{caption}
\usepackage{subcaption}
\usepackage[dvipsnames]{xcolor}

\hypersetup{
	colorlinks=true,
	linkcolor=blue,
	citecolor=blue,
	urlcolor=blue,
}

\newcommand{\AdS}{\text{AdS}}
\newcommand{\CFT}{\text{CFT}}
\newcommand{\BTZ}{\text{BTZ}}
\newcommand{\RT}{\text{RT}}

\begin{document}
	\begin{flushright}
		YITP-SB-2025-18
	\end{flushright}
	\begin{center}  
		
		\vskip 2cm 

        \begin{center}
  \LARGE\textbf{Aspects of holographic entanglement using physics-informed-neural-networks}
\end{center}
        
		\vskip 1cm
		
		\renewcommand{\thefootnote}{\fnsymbol{footnote}}
		
		\centerline{
			{\large  Anirudh Deb} \footnote{anirudh.deb@stonybrook.edu} and {\large  Yaman Sanghavi} \footnote{yaman.sanghavi@stonybrook.edu}}
		
		\vspace{1cm}
		\centerline{{\it  C.~N.~Yang Institute for Theoretical Physics,  Stony Brook University, Stony Brook,}}
		\centerline{{\it NY 11794-3840, USA}}
		
		\vspace{1cm}
		
	\end{center}
	
	\vskip 0.3 cm

	\begin{abstract}
		We implement physics-informed-neural-networks (PINNs) to compute holographic entanglement entropy and entanglement wedge cross section. This technique allows us to compute these quantities for arbitrary shapes of the subregions in any asymptotically AdS metric. We test our computations against some known results and further demonstrate the utility of PINNs in examples, where it is not straightforward to perform such computations.
	\end{abstract}
	
	\newpage
	
	\tableofcontents
	
	\section{Introduction and summary}
	We study the holographic entanglement entropy (HEE) \cite{Ryu:2006bv,Ryu:2006ef} (see \cite{Nishioka:2018khk, Headrick:2019eth, VanRaamsdonk:2016exw, Casini:2009sr, Calabrese:2009qy, Nishioka:2009un, Witten:2018zxz, Ryu:2006ef,Takayanagi:2025ula} for reviews on entanglement entropy in field theory and holography) and holographic entanglement of purification, using \textit{physics-informed neural networks} (PINNs) \cite{raissi2017physics,raissi2019physics}.  The holographic entanglement of purification corresponds to entanglement wedge cross section (EWCS) in the bulk dual \cite{Umemoto_2018, Nguyen:2017yqw} (see also \cite{Jokela:2019ebz,BabaeiVelni:2019pkw,Jeong:2019xdr}). Computing the entanglement entropy and entanglement of purification in the CFT side are not easy problems, but on the holographic dual this translates into a geometric problem of finding extremal codimension-2 surfaces. We find these surfaces using \textit{neural networks} (NNs) which serve as universal function approximators. NNs are functions which are parametrized by a set of parameters, called weights and biases, which are adjusted in order to minimize a quantity, called the \textit{loss function}. If one uses physical conditions to define loss functions, the setup is called a physics-informed-neural-network. Since computing HEE and EWCS involve finding minimal area surfaces, this makes them a straightforward target for application of PINNs, where the Euler-Lagrange equation for the minimal surfaces along with relevant boundary conditions can be treated as a loss function to train the network. Alternatively, one can also use the area of the surface (with some boundary conditions) as the loss function itself, but based on our implementations and experiments with networks, defining the loss function using the differential equations appear to be more efficient in this regard. 
	
	There have been implementations of machine learning to holographic bulk reconstruction, computing Reyni entropies in lattice QFTs, etc (see for example \cite{Ahn:2024jkk, Lam:2021ugb, Park:2022fqy, Bulgarelli:2024yrz,Ahn:2024gjf,Ahn:2025tjp}). Here, we do a more straightforward study where given a metric, we find the HEE and EWCS for specified subregions.  The problem of finding minimal area surfaces or geodesics is a rather old one  and there also exist tools to study these problems. \textit{Surface Evolver}  \cite{em/1048709050} and a method by Chopp \cite{Chopp:1991uf} are examples of such tools. PINNs have also been used to determined minimal surfaces in Euclidean spaces. In contrast to the standard numerical techniques, which approximate the surface by triangulation, NNs approximate the surface with a smooth differentiable function. In situations with some symmetry or some other simplification, exact analytic results are obtainable, but determining minimal surfaces in a very general setup with no apparent symmetry is a nontrivial problem. In particular, the computation of EWCS is a constrained minimization problem where one needs to determine the minimal area surface such that the boundary of the surface is constrained to the Ryu-Takayanagi (RT) surface for two disconnected regions. This is a rather difficult computational problem for a general shape in arbitrary spacetime metric and we argue that PINNs provide an easy to use method for such computations. Although we restrict to asymptotically AdS metrics, the techniques of this paper are applicable to any spacetime metric. 
    
    A question of interest is that for a fixed area of entangling region, which shape maximizes the entanglement entropy. The shape dependence of HEE has been studied previously in several papers
	\cite{Astaneh:2014uba, Allais_2015, Fonda:2015nma, Carmi:2015dla, Lewkowycz:2018sgn, Bueno:2015xda, Klebanov:2012yf,Cavini:2019wyb,Seminara:2017hhh,Fonda:2014cca,Mezei:2014zla}. It has been shown that for a fixed area/perimeter of the entangling region, the sphere maximizes the entanglement entropy \cite{ Mezei:2014zla,Faulkner:2015csl,Bueno:2021fxb}. In the light of such studies on shape dependence, we illustrate the flexibility of the PINNs for arbitrary shapes and backgrounds in examples discussed in Sections \ref{sec:hee} and \ref{sec:ewcs}. We restrict ourselves to bulk spacetimes dimension 3 and 4, whose metrics are pure AdS and AdS-Schwarzchild. When the geometry is $\AdS_3$ or $\BTZ$, we verify the results of the NN against known analytic expressions for HEE and EWCS. In the case of $\AdS_4$ and $\AdS_4$-Schwarzchild geometries, we consider different setups. Section \ref{sec:ellipseinads4} discusses the HEE for ellipses of fixed perimeter and different aspect ratios in pure $\AdS_4$ and find agreement with the fact that the circle maximizes the entanglement entropy (see  \cite{Allais_2015} for the same analysis without NNs). In Section \ref{sec:circleinads4bh}, we compute HEE for a circle in $\AdS_4$-Schwarzchild metric for different horizon radius. Section \ref{twoellipseinads4} studies the shape dependence of EWCS for two identical ellipses with fixed shortest distance. Finally, Section \ref{sec:twocircles} computes the EWCS for two disjoint circles with radius $1$ and $2$, in $\AdS_4$-Schwarzchild geometry, for various horizon radius.  All in all, this neural network technique is complementary to already known computational techniques and various softwares available to find extremal surfaces. However, a numerical solver tailored specifically for the EWCS for generic shapes is not known to us. All the computations in this paper have been done using PyTorch library \cite{Paszke:2019xhz} in Python. We provide a brief summary of NNs and the architecture used in Appendix \ref{app:neuralnet}. 

    Although, we discuss only a handful of examples to show the utility of PINNs, there are several interesting directions to explore using this technique. One of the main advantages of the PINN setup in the study of holographic entanglement measures is that it allows one to study the quantities for arbitrary shapes of the subregions. In light of the previous studies of shape dependence of HEE, it would be interesting to construct a NN that finds the extremal surface with maximal area, given the area of the entangling surface is fixed. A similar question can be phrased for EWCS , where the distance between the subregions and the area of the entangling region is fixed. We have restricted ourselves to static geometries and it is natural to ask how can these techniques be used to study time dependent geometries. It maybe interesting to investigate phase transitions using these techniques. The computations in this paper do only $\AdS_3/\CFT_2$ and $\AdS_4/\CFT_3$, but this technique is general and can be used in higher dimensions. 
	
    \textbf{Organization of the paper:} The paper is organized as follows. In Section \ref{sec:hee} we present the results of computing HEE using the network. In Section \ref{sec:ewcs} we discuss the holographic entanglement of purification and its computation in the bulk by computing EWCS. Appendix \ref{app:neuralnet} provides the details about the neural network architecture used in our computations and also discuss possible numerical issues while using the code/technique.  Appendix \ref{app:diffeqn} collects the differential equations satisfied by minimal surfaces. 
	
	\textbf{Note:} While we were in the final stages of the project, papers discussing minimal surfaces in AdS using PINNs appeared \cite{Hashimoto:2025zmiupd,Hashimoto:2025upw}.
	
	\section{Holographic entanglement entropy}
	\label{sec:hee}
	The Ryu-Takayanagi proposal states that the entanglement entropy of a system $A$ for a CFT is given by the area of codimension-2 surface with minimal area (denoted by $\gamma_A$) in one higher dimensional AdS space such that the boundary of the surface equals the boundary of $A$ \cite{Ryu:2006ef}
	\begin{equation}
		S_A=\frac{\text{Area}(\gamma_A)}{4G_N}~.
	\end{equation}
	In this section, we test PINNs as a method of computing the HEE in some shapes and also do some simple computations for the analyzing the shape dependence, i.e., dependence of $\gamma_A$ on the shape of $A$. In the rest of the paper, we will denote the $\text{Area}(\gamma_A)$ by simply $\mathcal{A}$. 

    Before moving onto examples, let us give a brief picture of what the neural network (NN) is doing. For concreteness, let us consider the case of case of geodesics in $\AdS_3$. At a given constant time slice, the metric in Poincare coordinates is $ds^2=(dx^2+dz^2)/z^2$. Let us choose the interval to be $x\in \left[-\frac{l}{2},\frac{l}{2}\right]$ at $z=0$. The NN takes a parameter $u\in[0,1]$ and outputs coordinates $(x(u),z(u))$, which upon training the NN, represent the coordinates of the RT surface (geodesics). Before training, these points will generate some arbitrary curve in $x$-$z$ plane. To train the NN, we first choose a set $S$ of discrete points in the interval $[0,1]$, i.e., the input domain of NN. We then choose the loss function as sum of two terms, i.e., one coming from the bulk differential equation \eqref{eq:bulkEOM} for the minimal surfaces and one coming from imposing the boundary condition. For the bulk loss function, we take it to be the sum of squares of residuals of the differential equation evaluated at $S$. For the boundary loss function, we take it to be the sum of squares of difference between NN's prediction of boundary points and the actual desired boundary points, i.e. $\left(-\frac{l}{2},0 \right)$ and $(\frac{l}{2},0)$ (see appendix \ref{app:neuralnet} for more details). During training, the loss decreases and the curve starts approximating the geodesic anchored at $\pm l/2$ (see Figure \ref{fig:geodesicsbh}). 
    
    A similar exercise is performed for a 2-dimensional region in $\AdS_4$ where the metric for a constant time slice, in Poincare coordinates, is $ds^2=(dx^2+dy^2 + dz^2)/z^2$. Consider a simply connected region at $z=0$. In order to find the RT surface using NN, we first define a disk $\mathbb{B}^2$ as the domain with points being labeled as $(u_1,u_2) \in \mathbb{B}^2$. The NN maps $(u_1,u_2)$ to coordinates $\left(x(u_1,u_2),y(u_1,u_2),z(u_1,u_2)\right)$, which will be eventually trained to be the points of the RT surface. Before training, NN will map $\mathbb{B}^2$ to an arbitrary shape homeomorphic to $\mathbb{B}^2$. To train it, we first discretize the domain. The bulk and boundary loss function are defined in a similar way as for $\AdS_3$ (see appendix \ref{app:neuralnet} for details). Upon training, the loss function decreases and the network approximates the minimal surface better and better. In later sections, we will consider disjoint union of simply connected regions at $z=0$ and in order to find the RT surface, the domain needs to be appropriately chosen so that it is homeomorphic to the RT surface.
    
    \begin{figure}
			\centering
			\vspace{-0.3in}\includegraphics[width=0.5\linewidth]{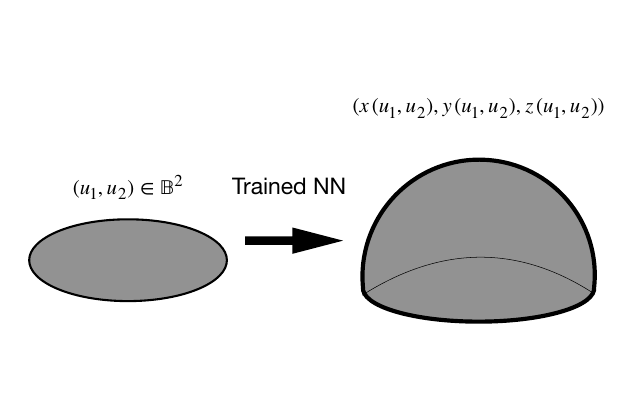}
            \vspace{-0.3in}
			\caption{A schematic illustration of a disk $\mathbb{B}^2$ being mapped to an RT surface via a neural network after training.}
			\label{fig:btzgoedesicpic}
	\end{figure}
    
	\subsection{$\AdS_3/\CFT_2$}
	\label{sec:HEEAdS3}
	
		\begin{figure}[t]
		\centering
		\begin{subfigure}[t]{0.4\linewidth}
			\centering
			\includegraphics[width=\linewidth]{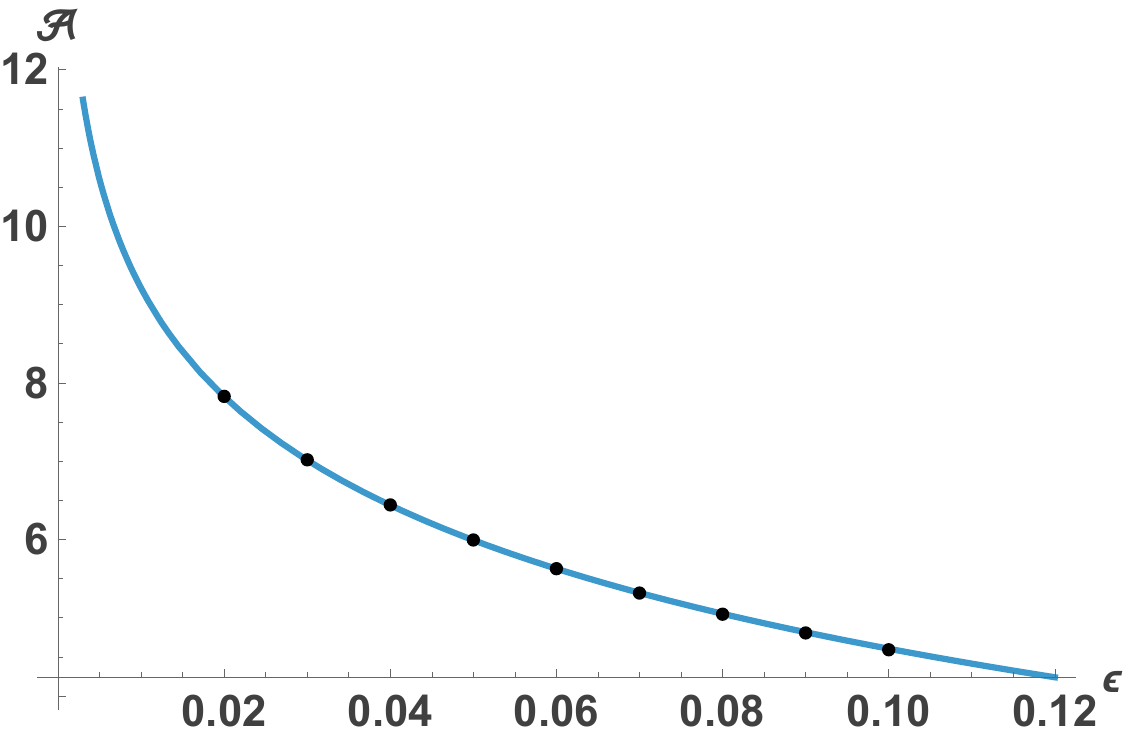}
			\caption{$\mathcal{A}$ vs $\epsilon$ for interval length $l$=1 in pure $\AdS_3$. The blue line displays the analytic result and the black dots are obtained using NN.}
			\label{fig:adsgeodesics}
		\end{subfigure}
		\hspace{1in}
		\begin{subfigure}[t]{0.4\linewidth}
			\centering
			\includegraphics[width=\linewidth]{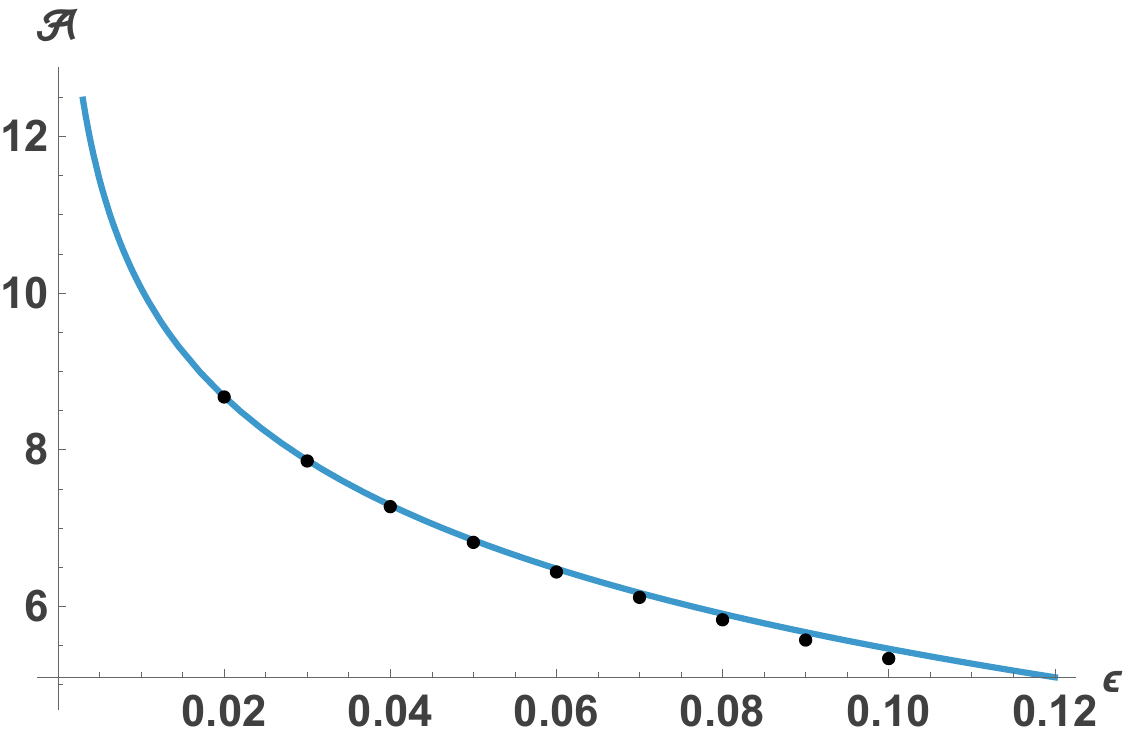}
			\caption{$\mathcal{A}$ vs $\epsilon$ for interval length $l$=1 in BTZ for $z_H=0.3$. The blue line displays the analytic result and the black dots are obtained using NN.} 
			\label{fig:btzgoedesic}
		\end{subfigure}
        \caption{}
	\end{figure}
	
\begin{figure}[t]
		
		\centering
		\includegraphics[width=0.7\linewidth]{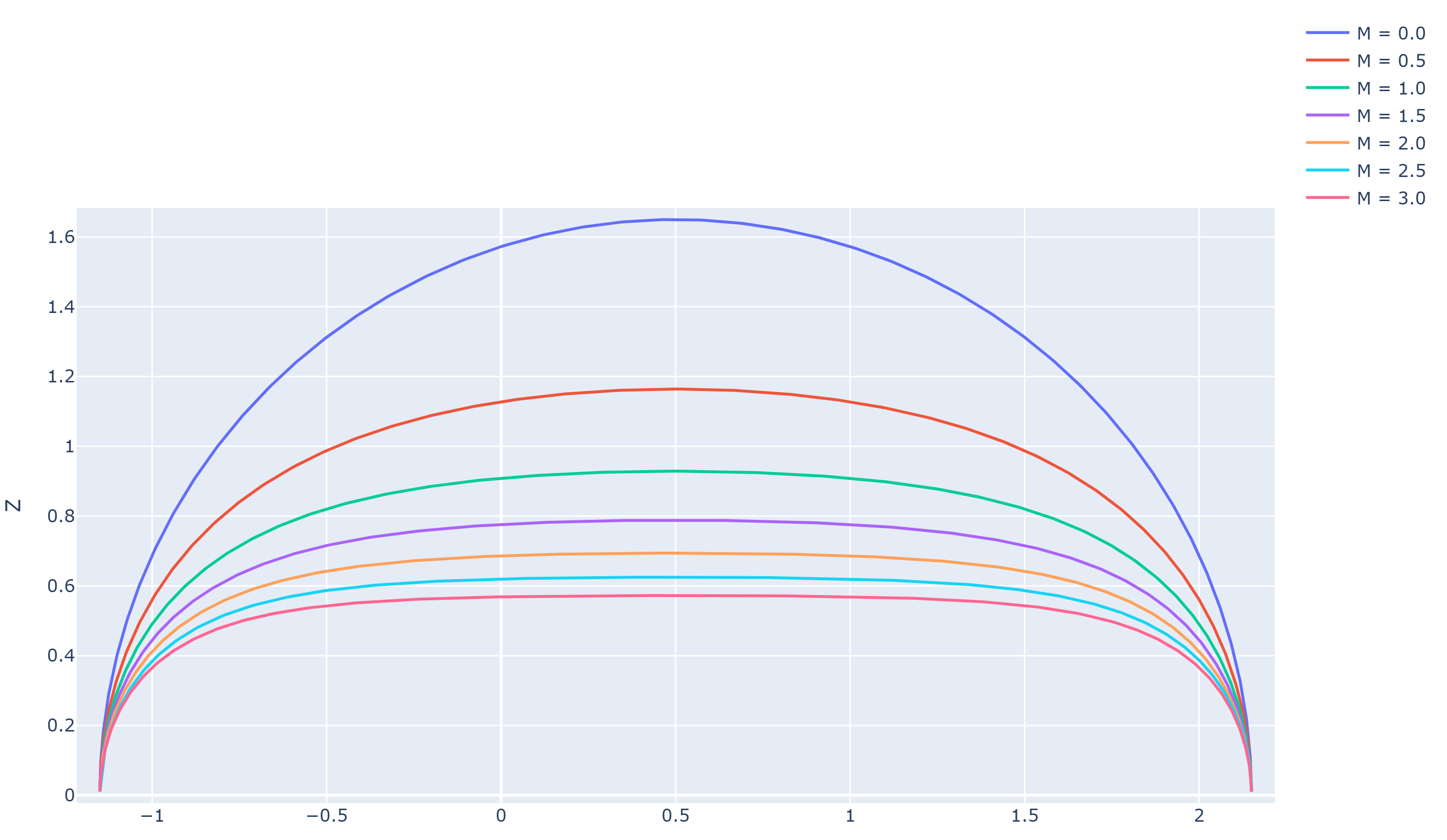}
		\caption{Geodesics in $\BTZ$ black hole geometry for different $M$ with fixed endpoints.}
		\label{fig:geodesicsbh}
	\end{figure}

	In this section, we consider the simplest application of the neural network by computing HEE when the dual geometry is $\AdS_3$ or $\BTZ$. The BTZ black hole has the following metric
	\begin{equation}
	\label{eq:btzmetric}
		ds^2=\frac{-\left(1-z^2/z_H^2\right)dt^2+\frac{dz^2}{\left(1-z^2/z_H^2\right)}+dx^2}{z^2}~,
	\end{equation}
	with $z_H$ the location of the horizon and $z_H=\infty$ is the case of the pure $\AdS_3$ (We will sometimes use $M=\frac{1}{z_H^2}$ instead of $z_H$). For pure $\AdS_3$, the minimal length geodesics are semicircles. The entanglement entropy for a system of length $l$ and using a cutoff $\epsilon$, is given as
	\begin{equation}
		S=\frac{1}{2G_N}\log\frac{l}{\epsilon}~.
	\end{equation}
	In Figure \ref{fig:adsgeodesics}, we plot $\mathcal{A}$ vs $\epsilon$ for $l=1$ and find agreement with the logarithmic divergence. We can perform the same computation for a BTZ black hole, where the entanglement entropy takes the form  
	\begin{equation}
		S=\frac{1}{2G_N}\log\left(\frac{\beta}{\pi \epsilon}\sinh\left(\frac{\pi l}{\beta}\right)\right)~,
	\end{equation}
	with $\beta=2\pi z_H$. Figure \ref{fig:btzgoedesic} shows divergence with $\epsilon$ for $z_H=0.3$. Figure \ref{fig:geodesicsbh} shows geodesics computed using PINN for various values of $M$.

	\subsection{$\AdS_4/\CFT_3$}
	\label{sec:HEEshapedep}
	Having tested the NN for geodesics, let us compute HEE for 2-dimensional subsystems. We study the HEE for ellipse in $\AdS_4$ and circular disk in $\AdS_4$-Schwarzschild black hole geometry. The metric is given as follows
	\begin{equation}
			\label{eq:btzzdt0}
		ds^2=\frac{-\left(1-z^3/z_H^3\right)dt^2+\frac{dz^2}{\left(1-z^3/z_H^3\right)}+dx^2+dy^2}{z^2}~,
	\end{equation}
	with $z_H$ denoting the position of the horizon and $z_H=\infty$ being the pure $\AdS_4$ case (we will occasionally use $M=\frac{1}{z_H^3}$ instead of $z_H$ ).  
	\subsubsection{Ellipse in $\AdS_4$}
    \label{sec:ellipseinads4}

    \begin{figure}[t]
		\centering
		\begin{subfigure}[t]{0.4\linewidth}
			\includegraphics[width=\linewidth]{{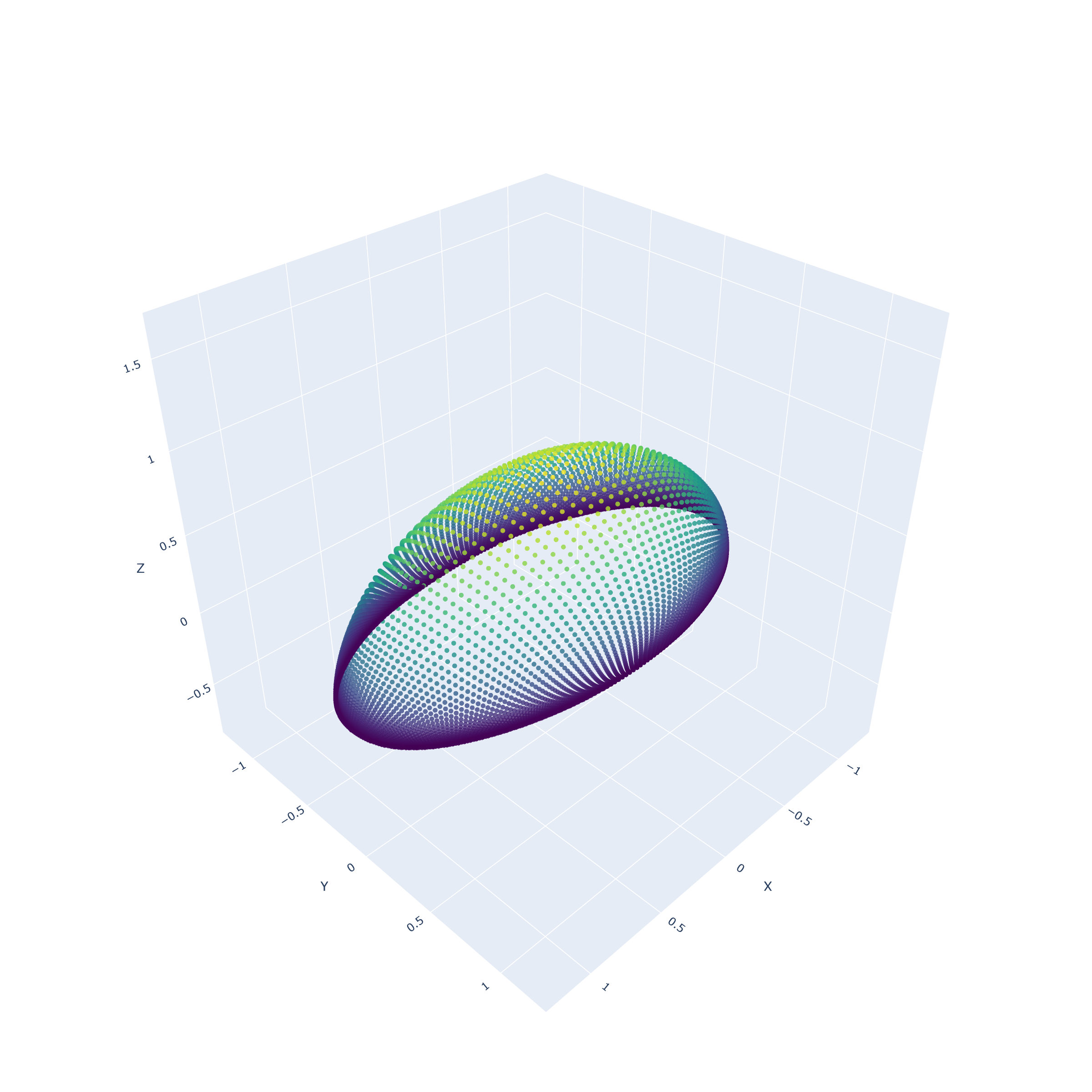}}
			\caption{RT surface in $\AdS_4$ for an entangling surface as an ellipse with $\alpha=0.5$.}
			\label{fig:EllipseinAdS4}
		\end{subfigure}
		\hspace{1in}
		\begin{subfigure}[t]{0.4\linewidth}
			\centering
			\includegraphics[width=\linewidth]{{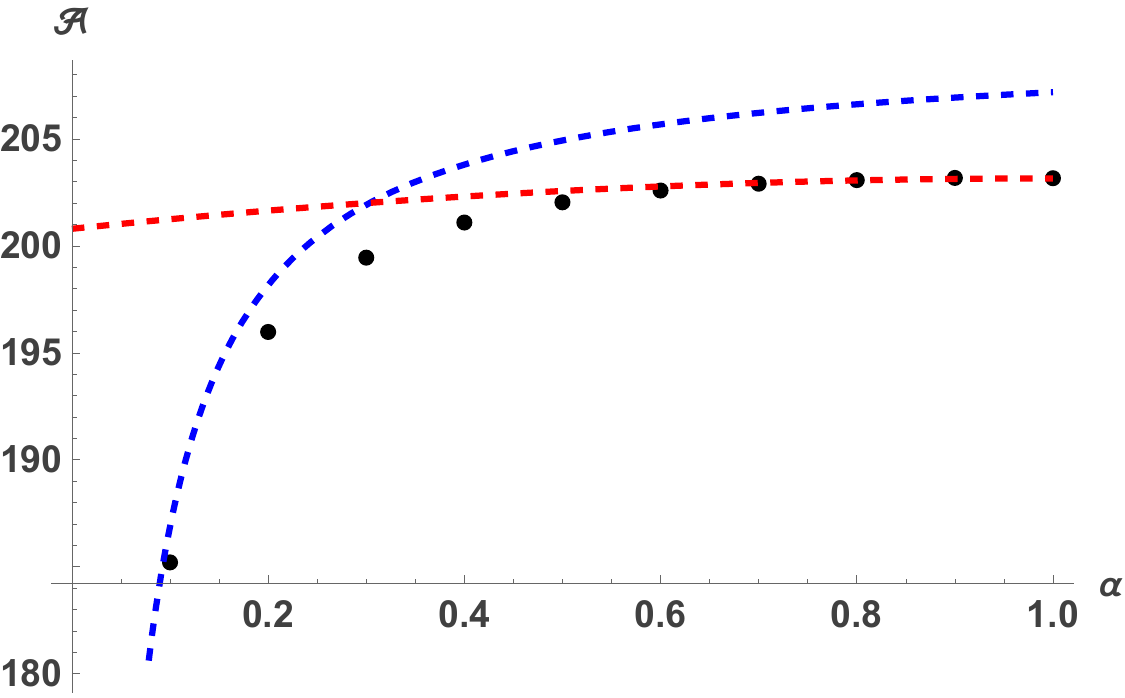}}
			\caption{HEE vs aspect ratio $\alpha$ for ellipse as the entangling surface. The black dots are the numerical values obtained using PINN. The blue line represents the inequality in  equation \eqref{eq:thinstripapprox}) and the red line is the approximation in equation \eqref{eq:alphaoneapprox}}
			\label{fig:HEEvsaspectratioAdS4}
		\end{subfigure}
		\caption{}
	\end{figure}
    
	Let us consider an ellipse in pure $\AdS$ metric
	\begin{equation}
		\frac{x^2}{a^2}+\frac{y^2}{b^2}=1~
	\end{equation}
	and analyze the dependence of HEE on the aspect ratio $\alpha=b/a$ for a fixed perimeter.  For a given $\alpha$ and perimeter $2\pi$, the semi-minor axes $b$ is given as
	\begin{equation}
		\label{eq:ellipseb}
		b=\frac{\pi}{\frac{2}{\alpha}E\left(1-\alpha^2\right)}~,
	\end{equation}
	where $E()$ is the elliptic integral of second kind.
	Example of a RT surface for $\alpha=0.5$ is displayed in figure \ref{fig:EllipseinAdS4}. Figure  \ref{fig:HEEvsaspectratioAdS4} displays $\mathcal{A}$ as a function of $\alpha$ for cutoff $\epsilon=0.03$. The is consistent with the theorem that circular disk maximizes the entanglement entropy over all entangling regions with the same perimeter \cite{ Mezei:2014zla,Faulkner:2015csl,Bueno:2021fxb}. For $\alpha$ close to $0$ and $1$, the results can be tested against the expressions in  \cite{Allais_2015}. For $\alpha\sim 0$
    \begin{equation}
    \label{eq:thinstripapprox}
        \mathcal{A}\leq \frac{2\pi}{\epsilon}-2\pi^2\left(\frac{\Gamma(\frac{3}{4})}{\Gamma(\frac{3}{4})}\right)^2\frac{1}{\alpha}~.
    \end{equation}
    For $\alpha\sim 1$, $\mathcal{A}$ is approximated by
    \begin{equation}
    \label{eq:alphaoneapprox}
        \mathcal{A}\sim\frac{2 \pi }{\epsilon}-2 \pi  \left(1+\frac{3}{8} (1-\alpha )^2\right)~.
    \end{equation}
   The blue dashed line in Figure \ref{fig:HEEvsaspectratioAdS4} represents the bound in equation \eqref{eq:thinstripapprox} and the red dashed line represents the approximation in equation \eqref{eq:alphaoneapprox}.

	\subsubsection{Circular Disk for bulk $\AdS_4$-Schwarzchild black hole geometry}
    \label{sec:circleinads4bh}
\begin{figure}[t]
		\centering
		\begin{subfigure}[t]{0.4\linewidth}
			\includegraphics[width=\linewidth]{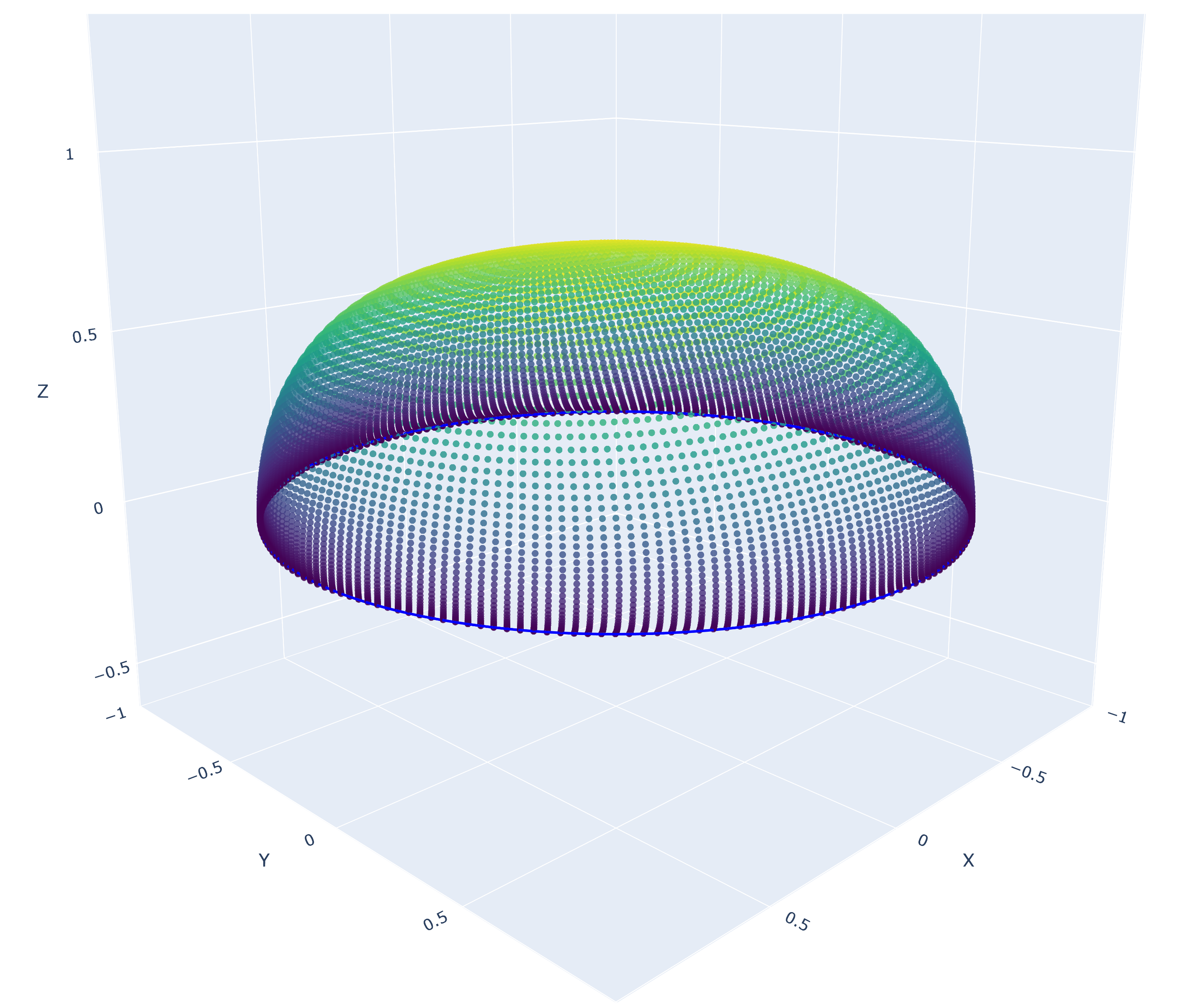}
			\caption{RT surface in AdS-Schwarzchild geometry with $M=2$, whose entangling surface is a unit circle.}
			\label{fig:circlepureAdS4}
		\end{subfigure}
		\hspace{1in}
		\begin{subfigure}[t]{0.4\linewidth}
			\includegraphics[width=\linewidth]{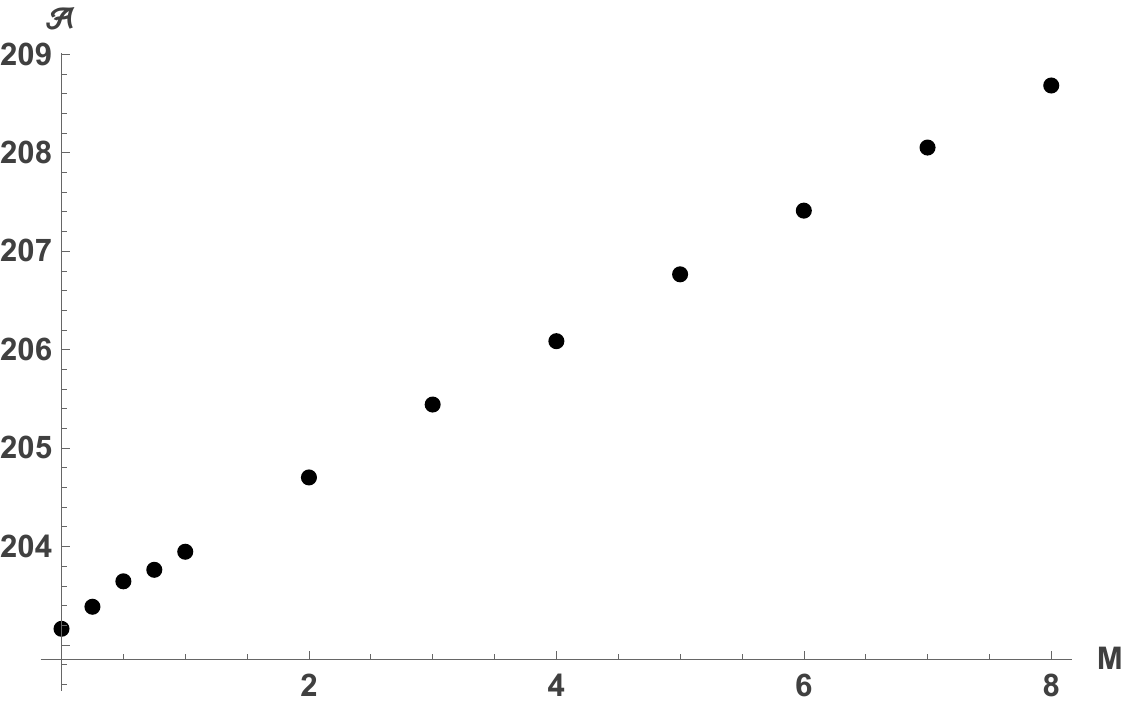}
			\caption{$\mathcal{A}$ vs $M$ for an RT surface, in $\AdS_4$-Schwarzchild black hole, whose entangling surface is a unit circle.}
			\label{fig:S_vs_r_AdS_4}
		\end{subfigure}
		\caption{}
	\end{figure}

	Next, consider a circular disk with radius $1$ in $\AdS_4$-Schwarzchild black hole metric with equation \eqref{eq:btzzdt0}. We do computations in terms of parameter $M=\frac{1}{z_H^3}$. Figure \ref{fig:circlepureAdS4} displays the RT surface for $M=2$ and \ref{fig:S_vs_r_AdS_4} displays the plot $\mathcal{A}$ vs $M$ for $\epsilon=0.03$. 

	\section{Entanglement wedge cross section}
	\label{sec:ewcs}

	\begin{figure}[t]
		\centering
		\includegraphics[]{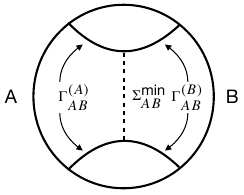}
		\caption{Entanglement wedge cross section.}
		\label{fig:defEWCS}
	\end{figure}
    
	Using the entanglement entropy, a very simple measure of correlation between two subsystems $A$ and $B$ can be introduced, which is called mutual information and is defined as
	\begin{equation}
		I(A,B)=S_A+S_B-S_{A\cap B}~.
	\end{equation}
	We use the notation $\mathcal{I}(A,B)=4G_N I(A,B)$. Even though mutual information is defined holographically, it is not a completely new quantity from the holographic point of view, as it can be defined in terms of entanglement entropy. Another correlation measure for a mixed state, called the \textit{entanglement of purification}, which was introduced in \cite{Terhal:2002riz} is new from the holographic point of view. This quantity is not a genuine entanglement measure because it is not always vanishing for separable states. Still it is a correlation measure that can be considered. Consider two subsystems $A$ and $B$, and consider $|\psi\rangle\in\mathcal{H}_{AA'}\otimes \mathcal{H}_{BB'}$. The entanglement of purification is defined as 
	\begin{equation}
		E_P(\rho_{AB})=\text{min}_{\rho_{AB}=\text{Tr}_{A'B'}|\psi\rangle\langle\psi|}S(\rho_{AA'})~.
	\end{equation}
	The holographic dual of this quantity was proposed to be equal to the entanglement wedge cross section $E_W(\rho_{AB})$ \cite{Umemoto_2018}. To define EWCS, consider two sub regions $A$ and $B$. Let $\Gamma_{AB}^{\text{min}}$ denote the RT surface (consider the choice of $A$ and $B$ such that the RT surface is the connected one). The entanglement wedge $M_{AB}$ is defined as the region surrounded by $A$, $B$ and $\Gamma_{AB}^{\text{min}}$ (see figure \ref{fig:defEWCS}). i.e.
	\begin{equation}
		\partial M_{AB}=A\cup B\cup \Gamma^{\text{min}}_{AB}~.
	\end{equation} 
	Now consider partition of $\Gamma_{AB}^{\text{min}}$ such that
	\begin{equation}
		\Gamma_{AB}^{\text{min}}=\Gamma_{AB}^{(A)}\cup \Gamma_{AB}^{(B)}
	\end{equation}
	and let $\Sigma_{AB}^{\text{min}}$ be the minimal area surface such that $\partial \Sigma_{AB}^{\text{min}}=\partial(A\cup\Gamma^{(A)}_{AB})=\partial(A\cup\Gamma^{(B)}_{AB})$ and $\Sigma_{AB}^{\text{min}}$ is homologous to $A\cup\Gamma^{(A)}_{AB}$ inside $M_{AB}$. Upto a factor of $\frac{1}{4G_N}$, the EWCS is defined as minimal value of the area of  $\Sigma_{AB}^{\text{min}}$ considered over all partitions of the entanglement wedge. i.e.
	\begin{equation}		E_W(\rho_{AB})=\text{min}_{\Gamma^{(A)}_{AB}\subset \Gamma^{\text{min}}_{AB}}\left[\frac{\text{Area}(\Sigma_{AB}^{\text{min}})}{4G_N}\right]~.
	\end{equation}
	In simple words, EWCS is proportional to the minimal area of all possible cross sections of the entanglement wedge. 		
	The EWCS is a finite quantity and does not diverge as the cutoff goes to zero. We will denote the area of the cross section by $\mathcal{A}_{\text{EWCS}}$ ($\mathcal{A}_{\text{EWCS}}=4G_NE_W$). 
    
    In this section, we first compute EWCS in the case of pure $\AdS_3$ and $\BTZ$ black hole metric. For the case of asymptotically $\AdS_4$ metrics, we compute it in the case of two disjoint ellipse with a fixed perimeter and separated by a fixed minimal distance. Finally, we also take two circular disk subregions, one with radius $1$ and the other with radius $2$ with fixed minimal distance in a $\AdS_4$-Schwarzchild black hole metric. (See for example \cite{Jokela:2019ebz,BabaeiVelni:2019pkw} for computations of EWCS in subregions with symmetric shapes like spheres, slabs and creases).  

    Let us first briefly discuss the neural network architecture for the case of EWCS in $\AdS_3$. We start with the a constant time slice in Poincare coordinates with the metric $(dx^2+dz^2)/z^2$. We consider two disjoint intervals $I_A = [x_{A_1},x_{A_2}]$ and $I_B = [x_{B_1},x_{B_2}]$ on the $x$-axis. The first step is to find the RT surface corresponding to the entanglement wedge which will be a pair of geodesics, say $C_1$ connecting $(x_{A_1},0)$ and $(x_{B_2},0)$ and $C_2$ connecting $(x_{A_2},0)$ and $(x_{B_1},0)$. See the blue curves in Figure \ref{fig:EWCSgenchoiceAdS3}. These can be found using two distinct NNs, i.e., say $\text{NN}_{\text{RT}_1}$ for $C_1$ and $\text{NN}_{\text{RT}_2}$ for $C_2$ in the same way described in \ref{sec:hee}. For EWCS, we consider another neural network $\text{NN}_{\text{EWCS}}$ which takes a parameter $v \in [0,1]$ as input and outputs two points $(x(v),z(v))$ which, upon training, will represent the coordinates of the bulk of EWCS. In addition, we need two more trainable parameters, say $p_1$ and $p_2$ which will be respectively mapped via $\text{NN}_{\text{RT}_1}$ and $\text{NN}_{\text{RT}_2}$ to points $b_1 \in C_1$ and $b_2 \in C_2$. Upon training, they are meant to serve as the boundary points of EWCS which live on $C_1$ and $C_2$. The loss function to train $\{\text{NN}_{\text{EWCS}}, p_1, p_2\}$ is sum of three terms. The first term is the bulk loss function coming from imposing the bulk differential equation. The second term is a boundary loss function which essentially imposes an orthogonality condition between EWCS and the RT surface (see Appendix \ref{app:diffeqn} for details). The last term is another boundary loss function which imposes that the distance between predicted boundary points of $\text{NN}_{\text{EWCS}}$ with $b_1$ and $b_2$ vanish.

    In the case of asymptotically $\AdS_4$, the neural network setup is a bit involved. We define three distinct neural networks. The first of these three networks which we refer as $\text{NN}_{\text{RT}}$ takes a cylindrical domain $S^1 \times I$ and maps it to the RT surface upon training. Specifically, a point on $(u_1,u_2)$ where $u_1 \in [0,2\pi]$ and $u_2 \in [0,1]$ is mapped to $(x(u_1,u_2),y(u_1,u_2),z(u_1,u_2))$. We choose the cylindrical domain because the desired RT surface is homeomorphic to a cylinder\footnote{In order to find the topology of the RT surface for two disjoint subregions $A$ and $B$, one needs to verify if the mutual information for a connected RT surface is positive. If it becomes, negative then the combined RT surface is disconnected, i.e., the union of individual RT surfaces for $A$ and $B$}. The loss function needed to train this is similar to that of section \ref{sec:hee}, i.e. bulk loss from the differential equation and the boundary loss imposing the surface is attached to the desired boundary.

    The remaining two networks are for creating the boundary and the bulk of EWCS. The second network, which we refer to as $\text{NN}_{\text{EWCS}_{bd}}$ takes a parameter $t \in S^1$ and maps it to the cylinder $S^1\times I$ defined above as the domain for the RT surface. It is then mapped via $\text{NN}_{\text{RT}}$ to a curve on the RT surface. Specifically, $t \in [0,2\pi]$ is first mapped to $(u_1(t),u_2(t))$ and then mapped to $\left(x(u_1(t),u_2(t)),y(u_1(t),u_2(t)),z(u_1(t),u_2(t))\right)$. Finally, the third network, referred to as $\text{NN}_{\text{EWCS}}$ takes a point $(v_1,v_2)$ on a disk $\mathbb{B}^2$ and maps it to $(x(v_1,v_2),y(v_1,v_2),z(v_1,v_2))$. The last two networks $\text{NN}_{\text{EWCS}_{bd}}$ and $\text{NN}_{\text{EWCS}}$ are trained simultaneously with a loss function that is sum of three terms. The first term is the bulk loss function imposing the bulk differential equation. The second term imposes the orthogonality condition between the EWCS and the RT surface. The last term imposes that the distance between the predicted boundary of $\text{NN}_{\text{EWCS}}$ and the curve mapped by $\text{NN}_{\text{EWCS}_{bd}}$ vanishes (see Appendix \ref{app:neuralnet} for more details). Figure \ref{fig:NNforEWCS} shows a schematic illustration of the above.

\begin{figure}[t]
    	\centering		
        \hspace{0.1in}
        \includegraphics[width=0.9\linewidth]{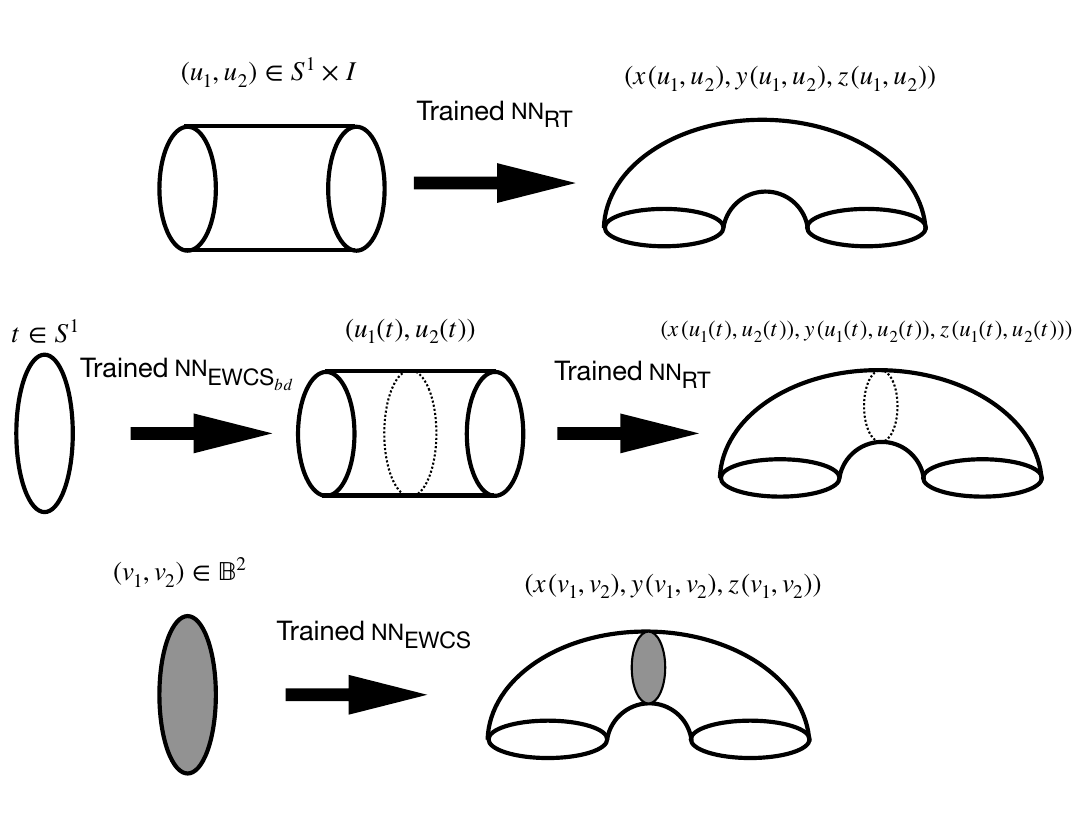}	
		\caption{A schematic illustration of the neural network architecture needed for generating an entanglement wedge and its minimal area cross section. It consists of three distinct neural networks. On the top, a network $\text{NN}_{\text{RT}}$ takes a cylinder domain $S^1 \times I$ to an RT surface. In the middle, an $S^1$ domain is mapped via another network $\text{NN}_{\text{EWCS}_{bd}}$ to a curve on the cylinder domain which is then mapped via $\text{NN}_{\text{RT}}$ to a curve on the RT surface which will serve as the boundary of the EWCS. At the bottom, $\mathbb{B}^2$ is mapped via the third network $\text{NN}_{\text{EWCS}}$ to the bulk of EWCS.}
        \label{fig:NNforEWCS}
	\end{figure}
    
	\subsection{$\AdS_3/\CFT_2$}
	\label{sec:EWCSAdS3}

    	\begin{figure}[t]
		\centering
        
		\begin{subfigure}[t]{0.4\linewidth}
			\centering
		\includegraphics[width=1\linewidth]{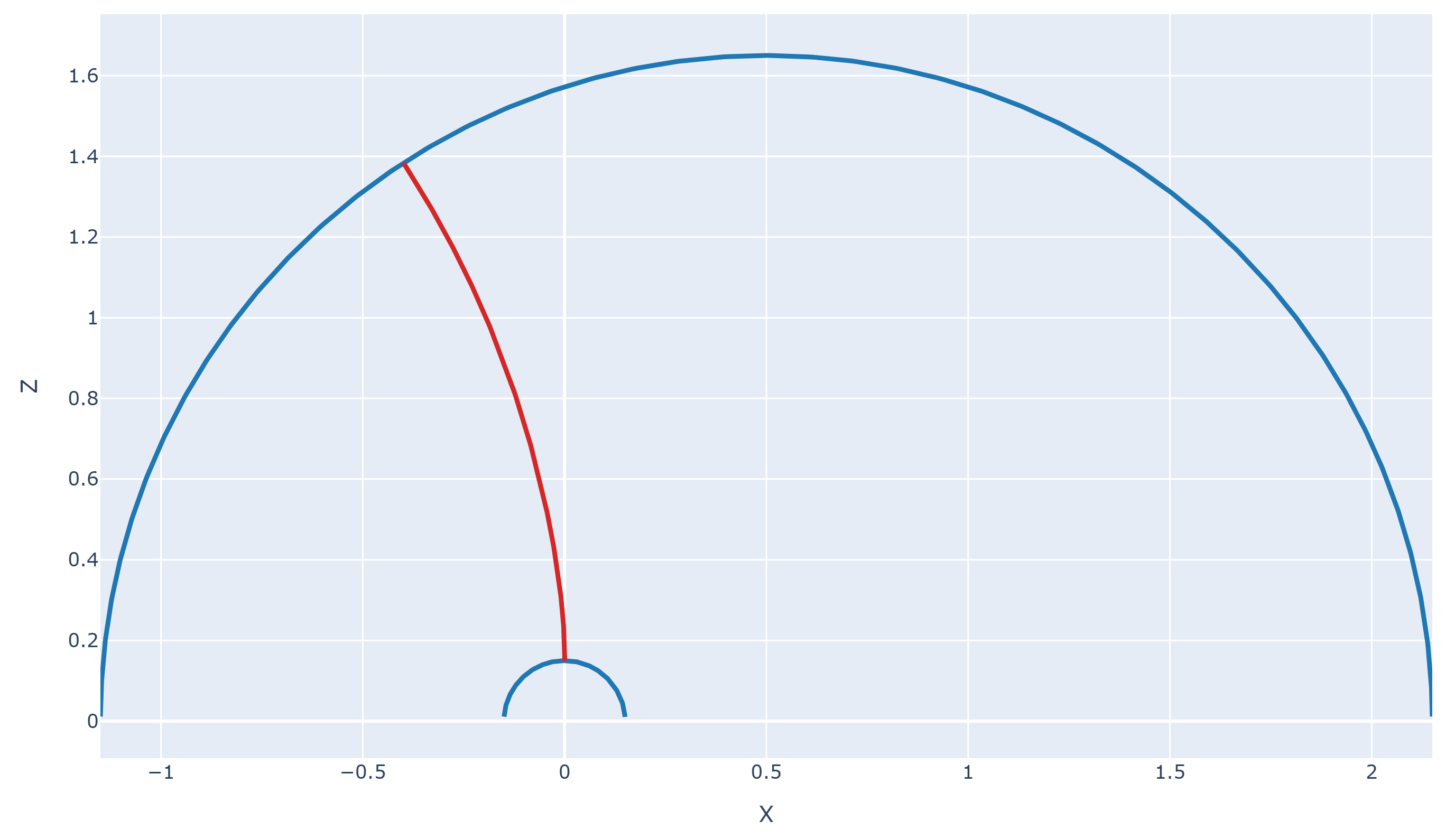}
			\caption{EWCS for $\AdS_3$ for intervals $[-1.15,-0.15]$ and $[0.15, 2.15]$.}
			\label{fig:EWCSgenchoiceAdS3}
		\end{subfigure}
        \hspace{1in}
       \begin{subfigure}[t]{0.4\linewidth}
			\centering
			\includegraphics[width=\linewidth]{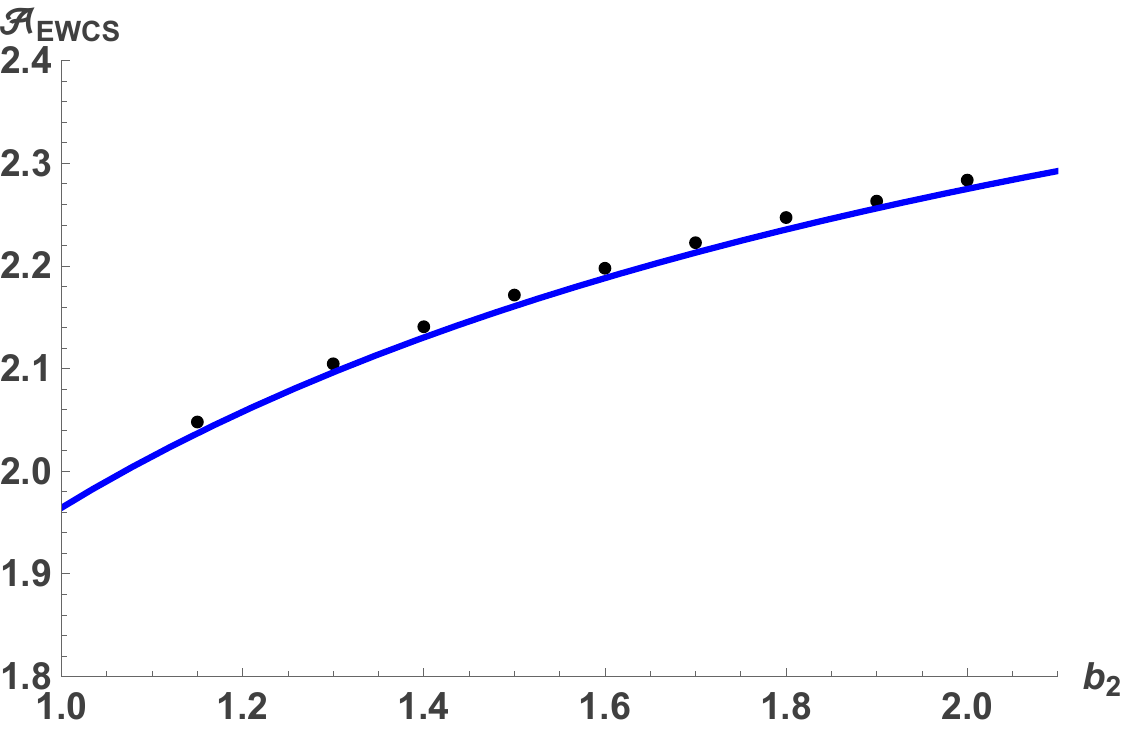}
			\caption{$\mathcal{A}_{\text{EWCS}}$ vs $b_2$ in $\AdS_3$ for intervals $[-1.15,-0.15]$ and $[0.15, 2.15]$.}
			\label{fig:EWCScircle}
		\end{subfigure}
        \caption{}
	\end{figure}
	
    \begin{figure}[t]
		\centering
         \begin{subfigure}[t]{0.4\linewidth}
			\centering
			\includegraphics[width=\linewidth]{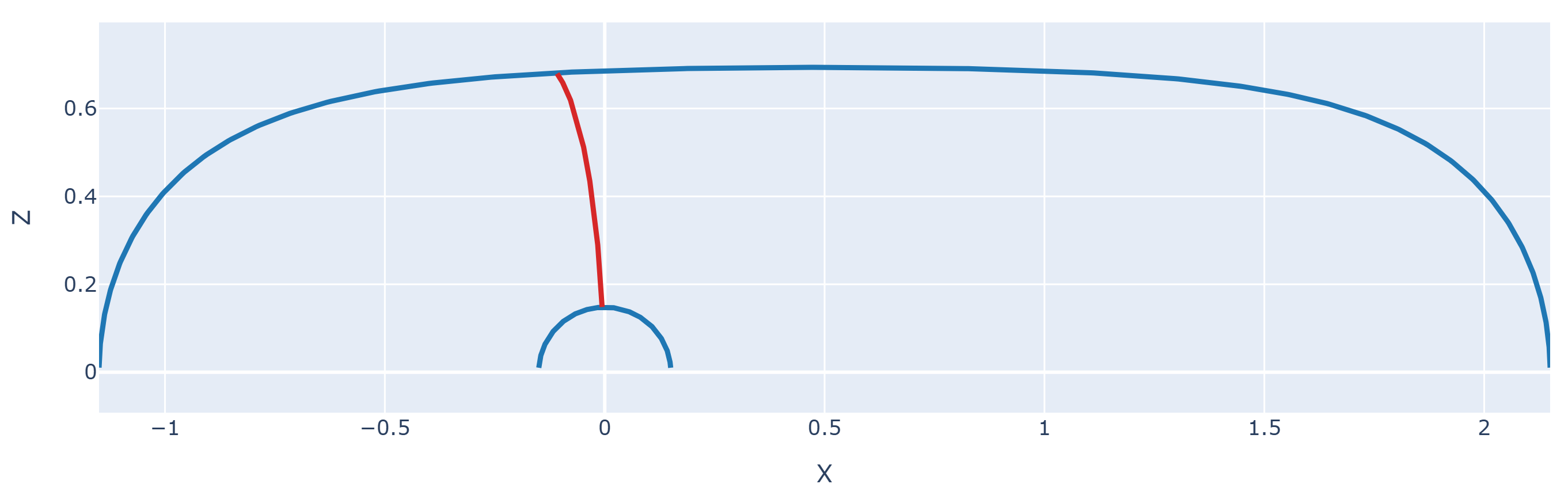}
			\caption{EWCS for $\BTZ$ for intervals $[-1.15,-0.15]$ and $[0.15, 2.15]$.}
			\label{fig:EWCSgenchoiceBTZ}
		\end{subfigure}
		\hspace{1in}
		\begin{subfigure}[t]{0.4\linewidth}
			\centering
			\includegraphics[width=\linewidth]{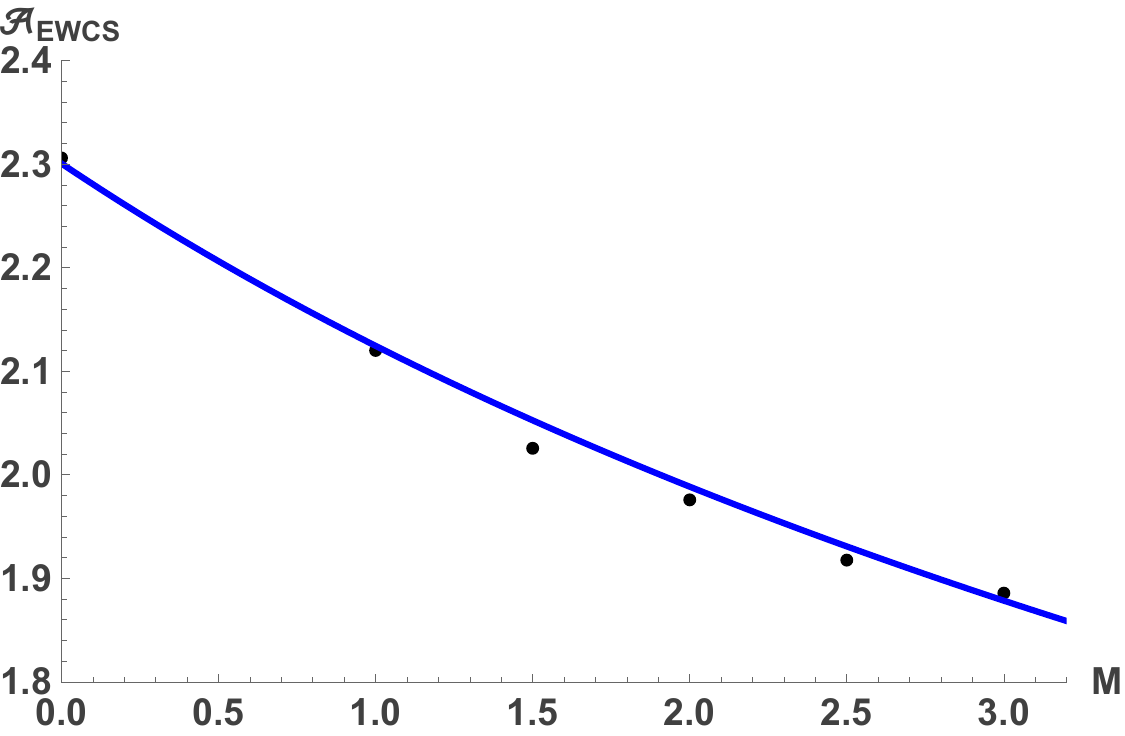}
			\caption{$\mathcal{A}_{\text{EWCS}}$ vs $M$ in $\BTZ$ for intervals $[-1.15,-0.15]$ and $[0.15, 2.15]$.}
			\label{fig:btzvsb1}
		\end{subfigure}
		\caption{}
	\end{figure}

	Let us first discuss the case of $\AdS_3/\CFT_2$. Consider two intervals $A=[a_1,a_2]$ and $B=[b_1,b_2]$. With choice of metric $(-dt^2+dz^2+dx^2)/z^2$, the corresponding RT surface are geodesics which are semicircles with center on the $x$-axis. A plot of EWCS for $a_1=-1.15$, $a_2=-0.15$, $b_1=0.15$ and $b_2=2.15$ is shown in Figure \ref{fig:EWCSgenchoiceAdS3}. The  \textcolor{blue}{blue} lines are the RT geodesics and the \textcolor{Red}{red} line is the EWCS geodesic. The value computed from the NN can be matched with the closed form expression \cite{Umemoto_2018}
	\begin{equation}
		E_W(\rho_{AB})=\frac{c}{6}\log\left(1+2z+2\sqrt{z(z+1)}\right)~,
	\end{equation}
	\begin{equation}
		z=\frac{(a_2-a_1)(b_2-b_1)}{(b_1-a_2)(b_2-a_1)}~.
	\end{equation}
	We fix the values of $a_1=-1.15,a_2=-0.15,b_1=0.15$ and plot $\mathcal{A}_{\text{EWCS}}$ in Figure \ref{fig:btzvsb1} for various $b_2$ and find agreement with the analytic expression.
	
	We perform the same exercises for BTZ metric (Eq. \eqref{eq:btzmetric}).
     Plot for the choice $a_1=-1.15$, $a_2=-0.15$, $b_1=0.15$,$b_2=2.15$ and $M=2$ is given in Figure \ref{fig:EWCSgenchoiceBTZ}. We also fix the values  $a_1=-1.15,a_2=-0.15,b_1=0.15$ and plot $\mathcal{A}_{\text{EWCS}}$ vs $b_1$ in Figure \ref{fig:btzvsb1}. We find agreement with the following closed form expression 
	\begin{equation}
		E_W(\rho_{AB})=\log\left(1+2\zeta+2\sqrt{\zeta(\zeta+1)}\right)~,
	\end{equation}
	where 
	\begin{equation}
		\label{eq:EWCSBTZ}
		\zeta=\frac{\sinh\left(\frac{\pi(a_2-a_1)}{\beta}\right)\sinh\left(\frac{\pi(b_2-b_1)}{\beta}\right)}{\sinh\left(\frac{\pi(b_1-a_2)}{\beta}\right)\sinh\left(\frac{\pi(b_2-a_1)}{\beta}\right)}~.
	\end{equation}

	\subsection{$\AdS_4/\CFT_3$}
	\label{sec:EWCSshapedep}

    \begin{figure}[t]
		\centering
		\begin{subfigure}[t]{0.4\linewidth}
			\centering
			\includegraphics[width=\linewidth]{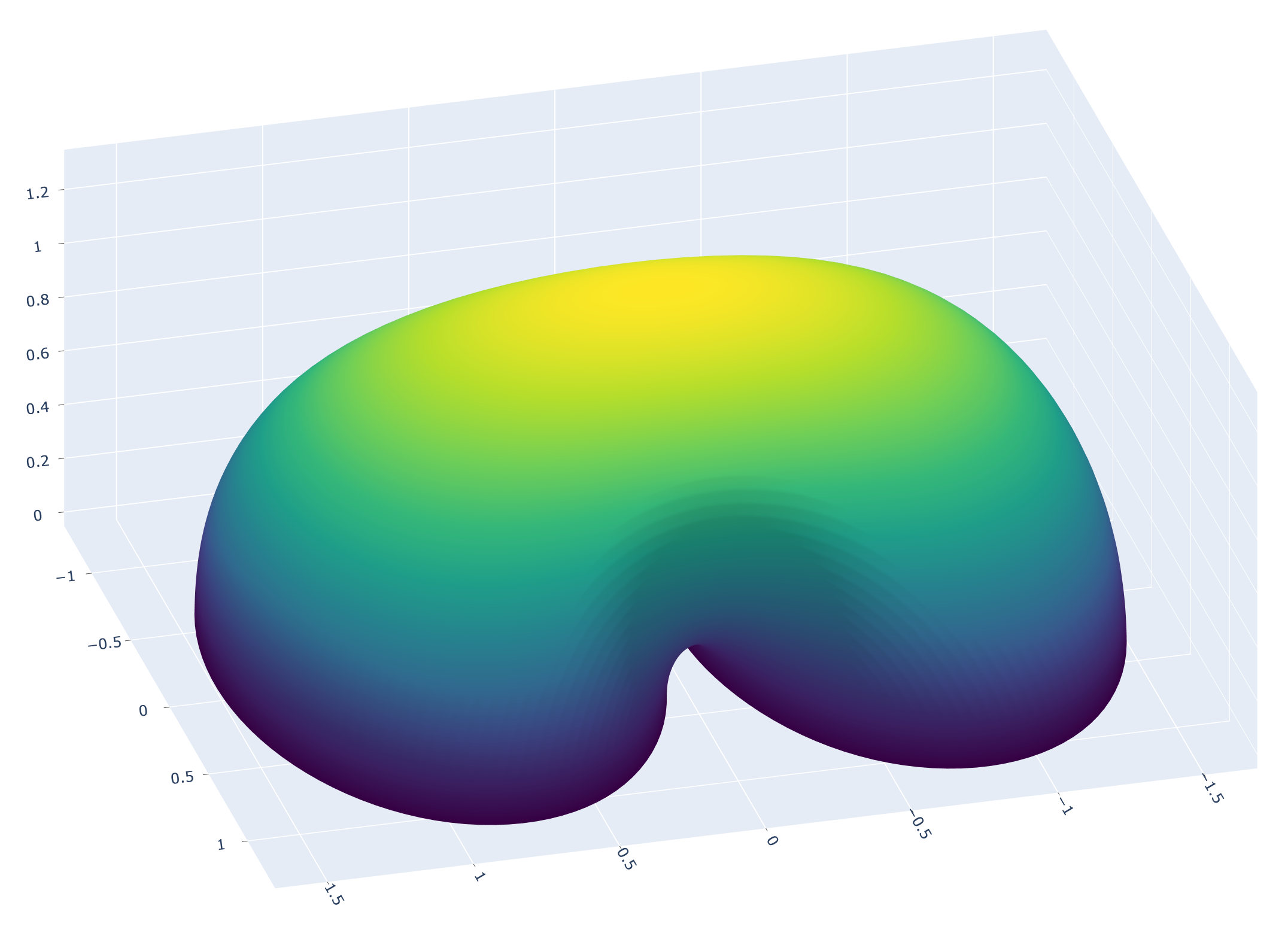}
			\caption{Connected RT surface  with boundary as the union of two congruent ellipses with perimeter $2\pi$, $\alpha=0.6$ and minimum distance $l=0.1$ between them.}
			\label{fig:EWCSellipseRT}
		\end{subfigure}
		\hspace{1in}
        \begin{subfigure}[t]{0.4\linewidth}
			\centering
			\includegraphics[width=\linewidth]{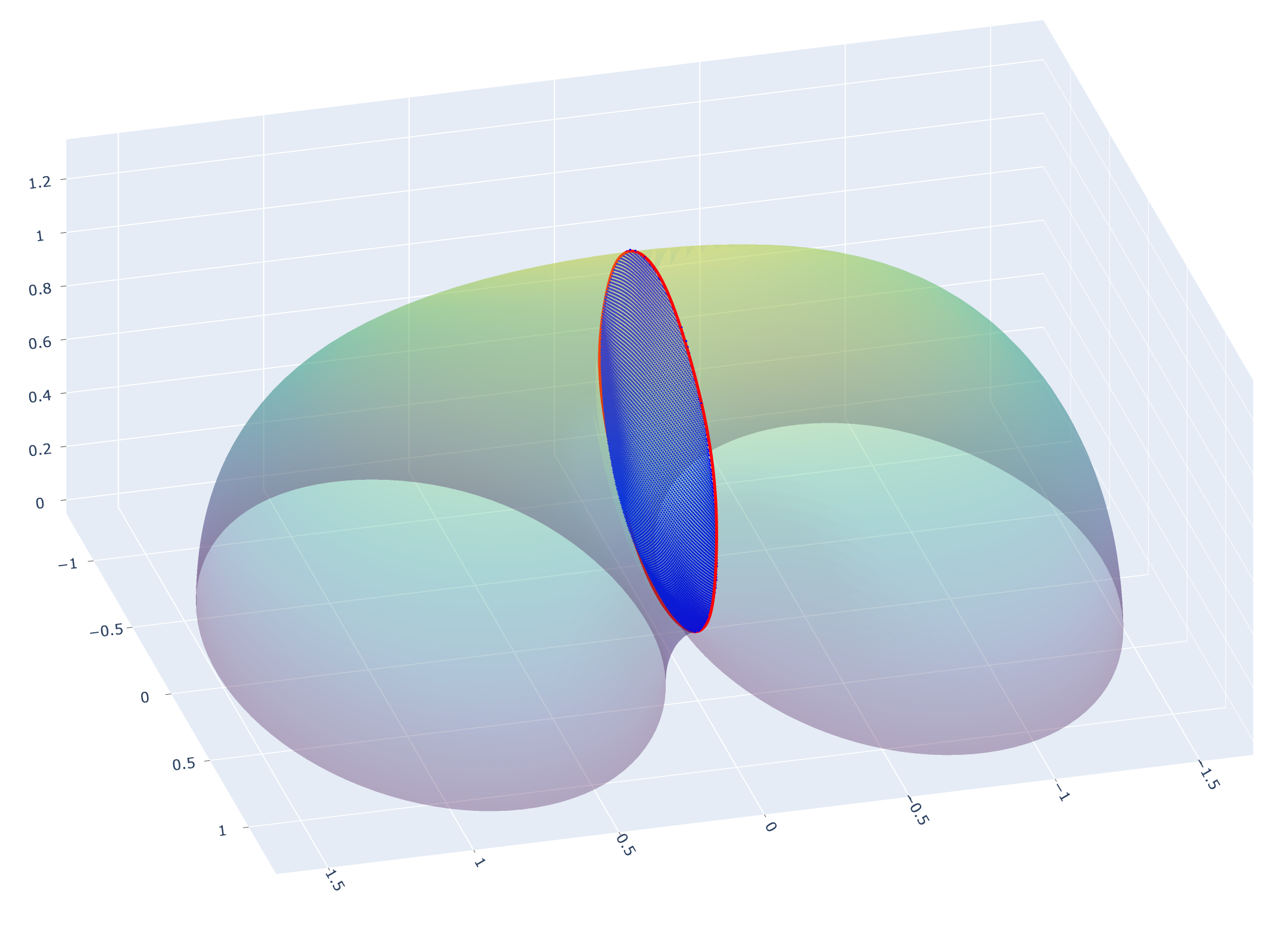}
			\caption{EWCS for an RT surface with boundary as the union of two congruent ellipses with perimeter $2\pi$, $\alpha=0.6$ and minimum distance $l=0.1$ between them. The RT surface is made transparent to show the EWCS (in blue).}
			\label{fig:EWCSellipse}
		\end{subfigure}
		\caption{}
	\end{figure}
        
        \begin{figure}[t]
			\centering
			\includegraphics[width=0.4\linewidth]{{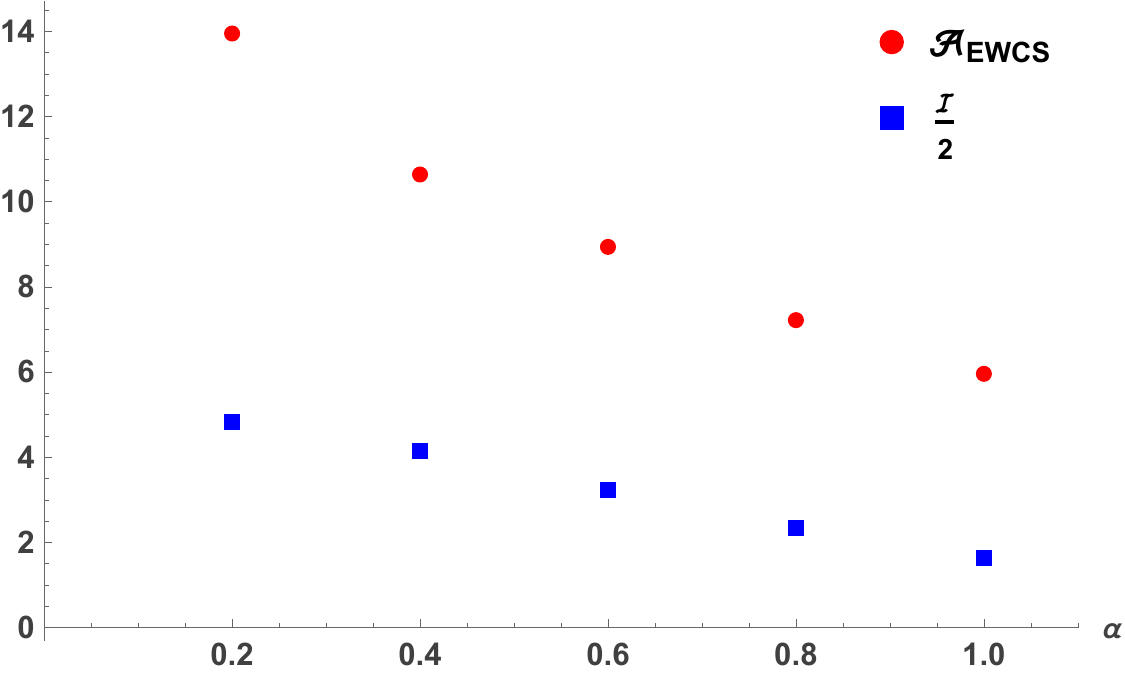}}
			\caption{EWCS and $I/2$ (where $I$ is mutual information) vs $\alpha$ for an RT surface, with boundary as the union of two congruent ellipses. The perimeter of each ellipse is $2\pi$ and minimum distance between them is $l=0.1$.}
			\label{fig:EWCSellipseplot}
	\end{figure}
    
	We can perform a shape dependence analysis for EWCS, similar to the one done for HEE surfaces in Section \ref{sec:HEEshapedep}. Let us define the separation between two non overlapping shapes to be the shortest distance between them. We consider various pairs of identical non overlapping ellipses with different aspect ratios keeping the perimeter fixed. The ellipses are placed such that their semi-major axes are on the same line and their separation is fixed. We choose this setup to study qualitatively the question as to which shapes with fixed perimeter and fixed separation maximize the EWCS. Next, we also illustrate the utility of PINNs by considering a slightly more complicated case of two disconnected circular subregions with radius $1$ and $2$, with a fixed separation in a black hole background. We study the dependence of EWCS for various values of $M$. 
	\subsubsection{Ellipse  in $\AdS_4$}
    \label{twoellipseinads4}
	Consider two ellipses with their centers separated by a distance $l+2b$ in pure $\AdS_4$. The equations being
	\begin{equation}
		\frac{x^2}{a^2}+\frac{\left(y\pm(\frac{l}{2}+b)\right)^2}{b^2}=1~.
	\end{equation}
	For a fixed perimeter, which we take to be $2\pi$, we vary aspect ratio $\alpha$ (see equation \eqref{eq:ellipseb}). We choose $l=0.1$ and $\alpha=0.2$, $0.4$, $0.6$, $0.8$ and $1.0$. For these choices a connected RT surface exists and has area less than those of the sum of two RT surfaces for individual ellipses. For choice $\alpha=0.6$, Figure \ref{fig:EWCSellipseRT} displays the connected RT surface and Figure \ref{fig:EWCSellipse} the entanglement wedge. Figure \ref{fig:EWCSellipseplot} displays half of the mutual information $\mathcal{I}/2$ and $\mathcal{A}_{\text{EWCS}}$ vs the aspect ratio $\alpha$.  We have not attempted to study this behaviour analytically, but both the quantities decrease as the shape becomes closer to the circle. This is consistent with the intuition that for a fixed distance, there will be less correlations when points on both the regions are farther apart from each other. The quantities also satisfy the inequality $E_W\geq I/2$.

	\subsubsection{Disconnected circles of different radii in black hole geometry}
    
    \label{sec:twocircles}
    	\begin{figure}
        \begin{subfigure}[t]{0.4\linewidth}
			\centering
	   \includegraphics[width=\linewidth]{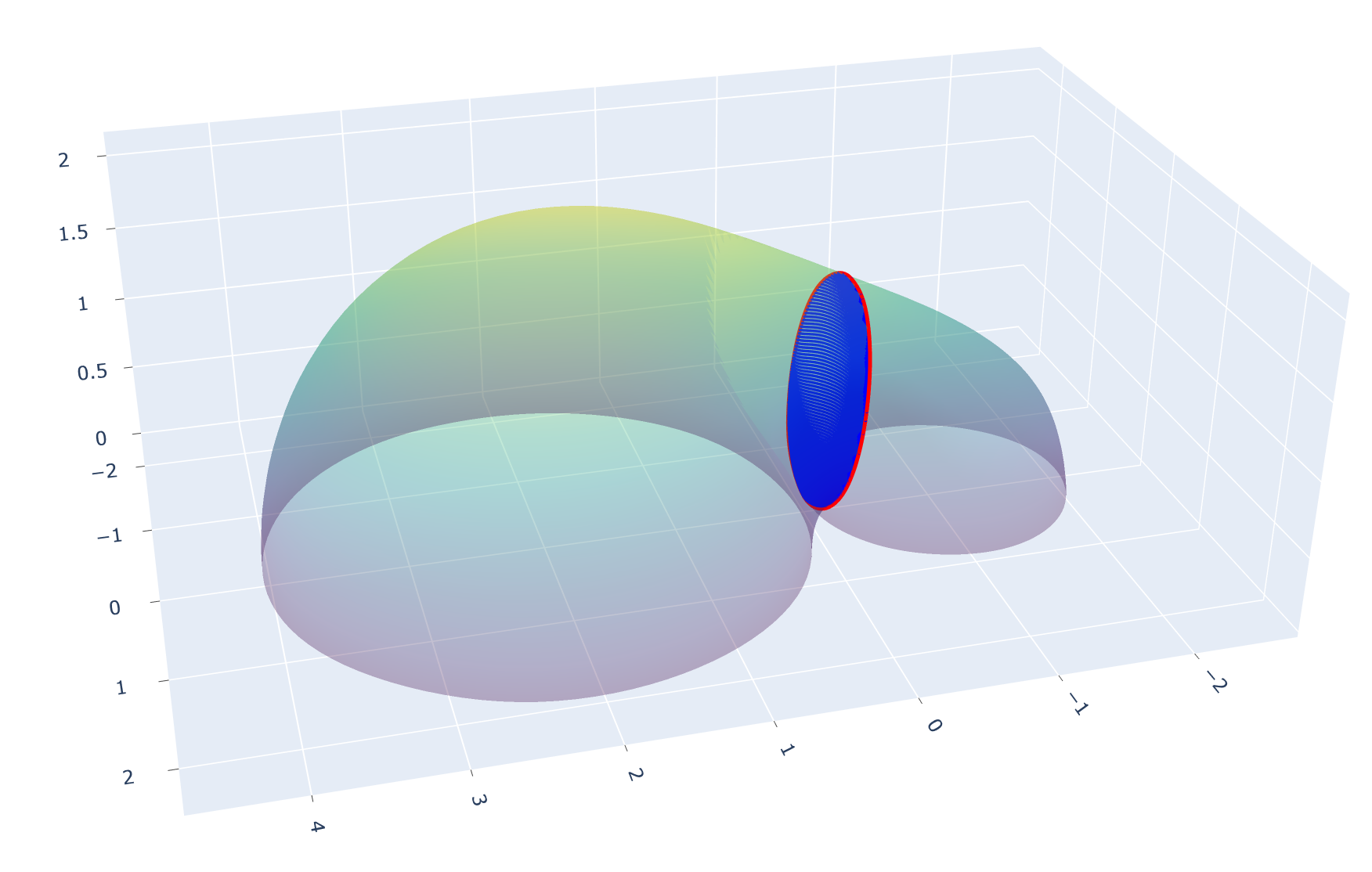}
			\caption{EWCS (in blue) and the connected RT surface for boundaries as union of two disjoint circles with radii $1$ and $2$, and minimum distance $0.1$, in pure $\AdS_4$.}
			\label{fig:EWCScirczh0}
		\end{subfigure}
        	\hspace{1in}
		\begin{subfigure}[t]{0.4\linewidth}
			\centering
			\includegraphics[width=\linewidth]{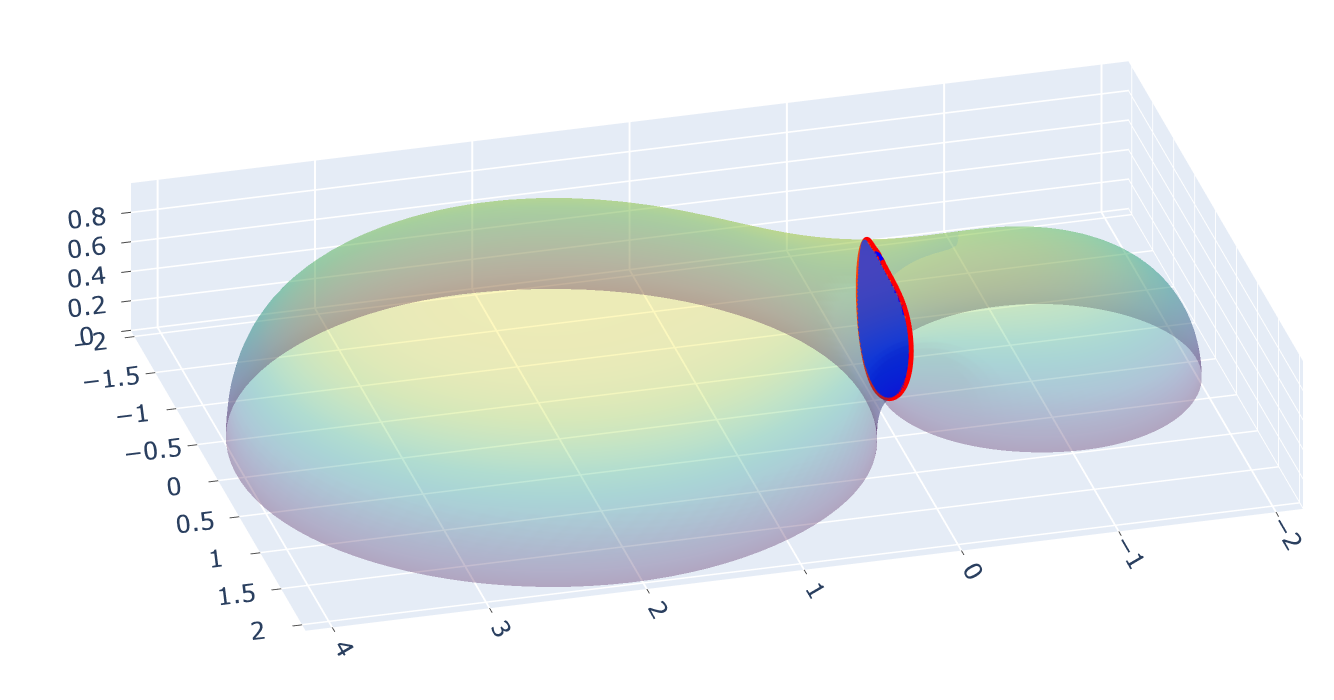}
			\caption{EWCS (in blue) and the connected RT surface with its boundaries as union of two circles with radii $1$ and $2$, and minimum distance $0.1$, in $\AdS_4$-Schwarzchild black hole with $M=1$.}
			\label{fig:EWCScirczhn0}
		\end{subfigure}	
        	\caption{}
    \end{figure}
\begin{figure}[t]
			\centering
    		\includegraphics[width=0.4\linewidth]{{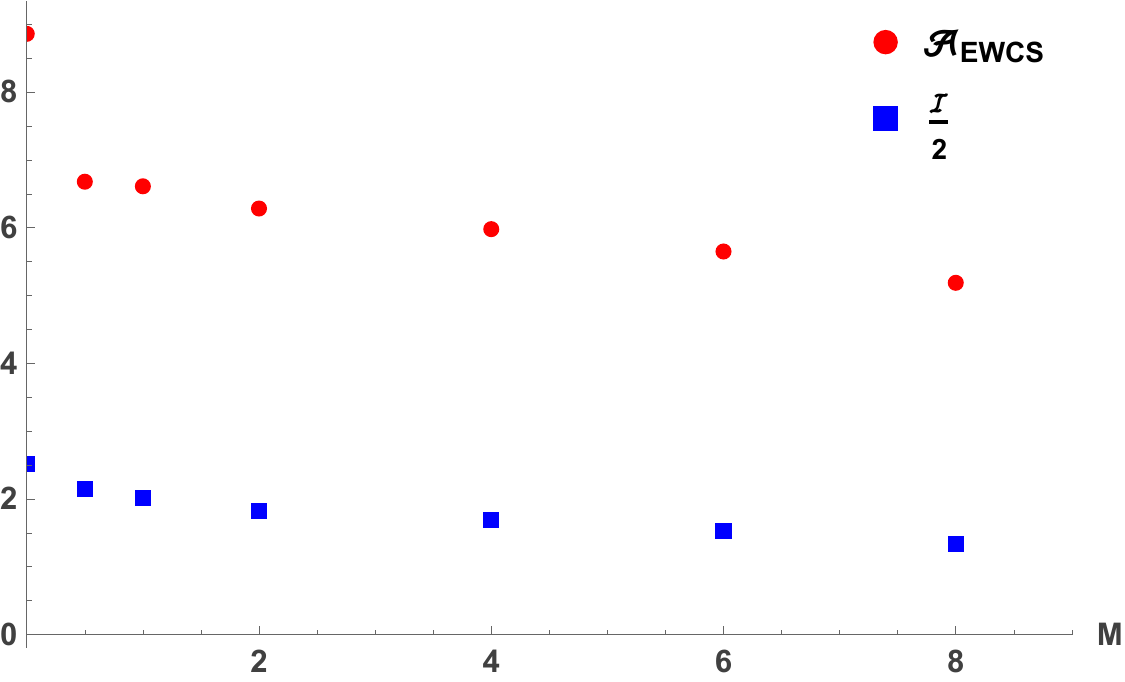}}
			\caption{EWCS and $I/2$ (where $I$ is mutual information) vs $M$ for an RT surface with its boundaries as union of two circles with radii $1$ and $2$, and minimum distance $0.1$, in $\AdS_4$-Schwarzchild black hole.}
			\label{fig:EWCScircleplot}
		\end{figure}

	In this section, we consider two circles with radius $1$ and $2$, with their centers separated by a fixed distance of $3.1$. We study this in $\AdS_4$-Schwarzchild background (see equation \eqref{eq:btzzdt0}). In this case, since the two subregions are not identical, the EWCS is not formed equidistant from the two subregions as in the case of Section \ref{sec:EWCSshapedep} and therefore it is not straightforward to compute this, even if the RT surface is provided. Figure \ref{fig:EWCScirczh0} and \ref{fig:EWCScirczhn0} display the RT surface and EWCS for $M=0$ and $M=1$ respectively. Figure \ref{fig:EWCScircleplot} displays the behaviour of half of the mutual information $\mathcal{I}/2$ and EWCS with $M$. The trend indicates that both quantities decrease as $M$ is increased. This is similar to the trend for two intervals in BTZ (see Figure \ref{fig:btzvsb1}). Also, the EWCS and mutual information satisfy the expected inequality $E_W\geq \frac{I}{2}$.

	\section*{Acknowledgments}
	We would like to thank for Jyotirmoy Bhattacharya for comments on the draft and valuable discussions. We would also like to thank Leonardo Rastelli and Shu-Heng Shao for useful discussions. The work of AD is supported in part by NSF grant PHY-2513893 and by the Simons Foundation grant 681267 (Simons Investigator Award). The work of YS was supported in part by the  Simons Collaboration on Ultra-Quantum Matter, which
is a grant from the Simons Foundation (651444, SHS).
	\appendix 
	\section{Neural network architecture and possible numerical issues}
	\label{app:neuralnet}
	In this appendix, we describe some basics of neural networks and explain the architecture used in this paper. See  \cite{Carleo:2019ptp,Halverson:2024hax,Mehta:2018dln,karniadakis2021physics,cuomo2022scientific,de2024numerical} for reviews on neural networks/machine learning for physicists and \cite{Schmidhuber:2014bpo,LeCun:2015pmr,bishop2006pattern,goodfellow2016deep} for general reviews, not targeted towards physics. We follow the lectures \cite{Halverson:2024hax} for covering basics.  A neural network takes $x\in\mathbb{R}^d$ as input and returns $\phi_\theta(x):\mathbb{R}^d\to\mathbb{R}$, where $\theta$ denotes the parameters of the network. The process of training adjusts the parameters such that a certain function $\mathcal{L}(\theta)$, called the loss function is minimized. The most straightforward approach is to optimize $\theta$ by performing gradient descent
	\begin{equation}
		\frac{d\theta_i}{dt}=-\nabla_{\theta_i}\mathcal{L}(\theta)~.
	\end{equation}
	There are various other algorithms to minimize the loss function. In our approach we use Adam optimizer \cite{Kingma:2014vow}. Once Adam optimizer reaches close to the minima, we use L-BFGS \cite{Liu:1989esw} to further refine the result.
	
	Now, let us discuss what a neural network consists of. It is a collection of neurons, where $\mu$-th neuron takes input as $x$ and returns $\sigma(w_\mu x+b_\mu)$.  $w_\mu$ is called weight and $b_\mu$ is called bias for $\mu$-th neuron. $\sigma$ is a non-polynomial linear activation function.  $\tanh()$ is an example of such an activation function. Example of a network with a single hidden layer of neurons is as follows
	\begin{equation}
		\label{eq:NNwb}
		\phi(x)=\sum_i^N\sum_j^d w^{(1)}_i \sigma(w_{ij}^{(0)}+b_i^{(0)})+b^{(1)}~,
	\end{equation}
	where  $d$ and $N$ depend on how big a network one would like to choose.The universal approximation theorem \cite{Cybenko:1989iql} states that any continuous function can be approximated by the above network. A more precise statement is that for a continuous function $f(x)$ and any $\epsilon>0$, there exists a NN of the form in equation \eqref{eq:NNwb} such that $|f(x)-\phi(x)|<\epsilon$. 
	
	For a $d$-dimensional subregion in $\CFT_{d+1}$, we choose a NN with $d$ inputs, $d+1$ outputs and $n$ layers. Let us denote the inputs as $u^{i=1,2,\dots d}$ and the outputs as $x^{\mu=1,2,\dots d+1}$. $x^\mu$ will denote the points in $\AdS_{d+2}$. The loss function is chosen to be the pointwise evaluation of the square of the differential equation satisfied by the of codimension-2 surface, plus constraints on the boundary of the hypersurface, which depend on whether we are computing the HEE or the EWCS. 
	\begin{figure}[t]
		\centering
        \hspace{-0.6in}
		\begin{subfigure}[t]{0.4\linewidth}
			\centering
\includegraphics[width=1.2\linewidth]{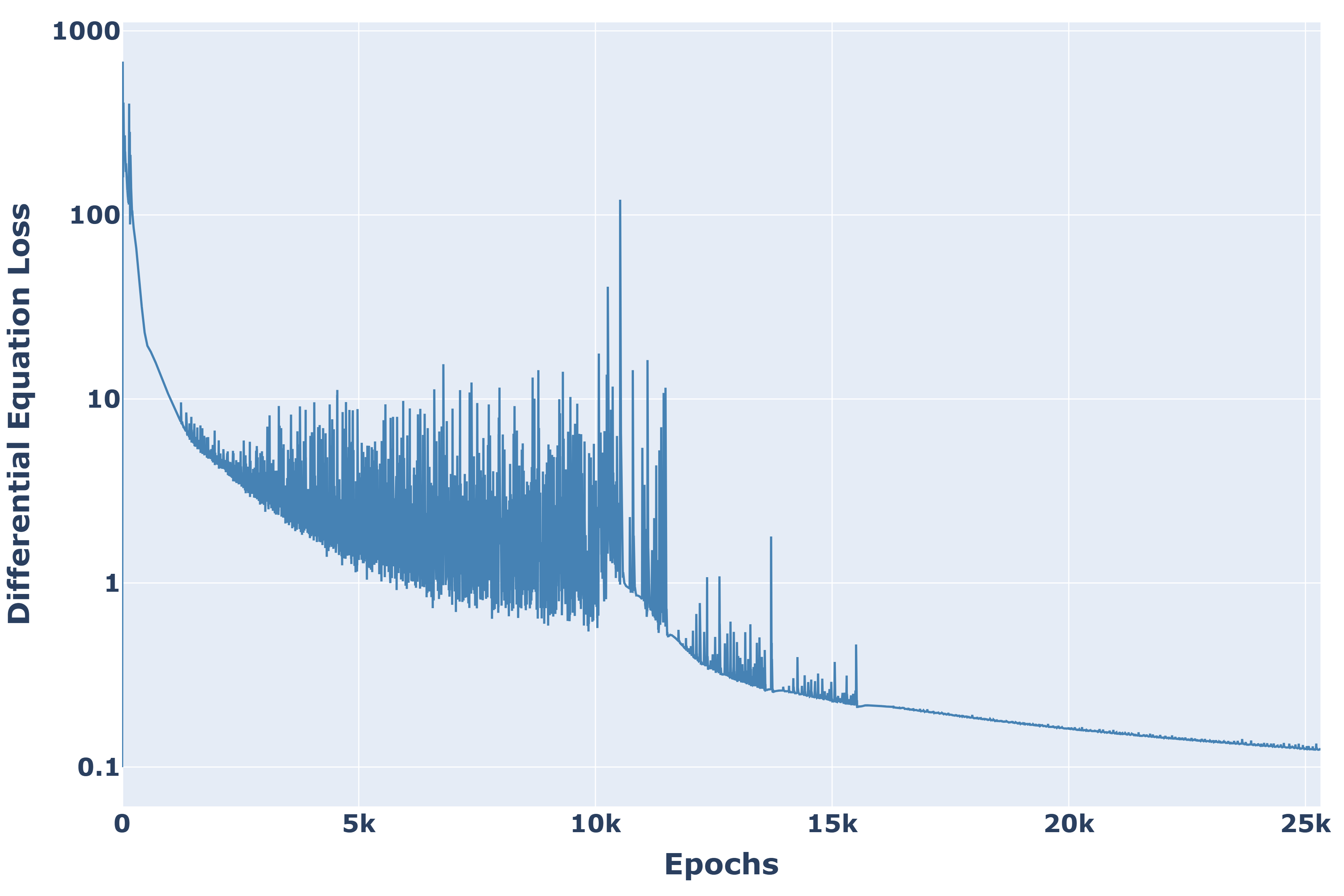}
			\label{}
		\end{subfigure}
        \hspace{0.7in}
       \begin{subfigure}[t]{0.4\linewidth}
			\centering
			\includegraphics[width=1.2\linewidth]{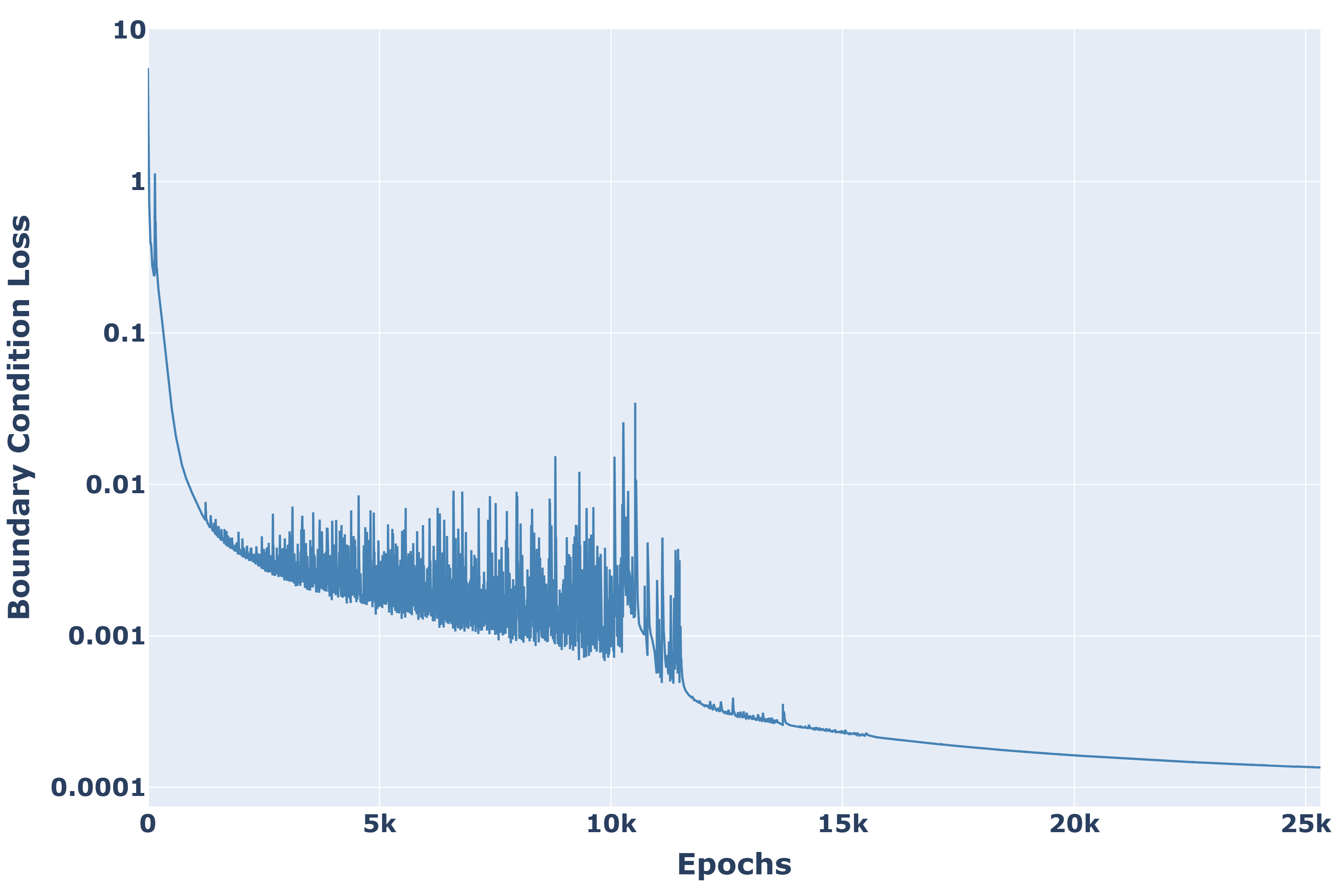}
			\label{fig:EWCScircle}
		\end{subfigure}
    \caption{Typical behavior of loss function over training epochs. This specific plot is for the RT surface for two unit circles with distance $ 2.1$ between their centers. On the left, we plot the differential equation loss and on the right, we plot boundary loss. For the first $10$K epochs, we used Adam optimizer with learning rate $=0.01$ and then for subsequent epochs, we used L-BFGS with learning rate $=0.1$.}

	\end{figure}

	Let us first discuss the case of HEE. Consider a compact domain $U$, which will parametrize the surface. We choose a set of $N$ discrete points in the domain and not on the boundary $u^i_{a=1,\dots N}\in U\backslash \partial U$. Another set of discrete points  $\mathtt{u}^i_{a=1,\dots , N_b}\in\partial U$ are chosen on the boundary of the domain. Let $x^\mu_{a}$ denote the output corresponding to $u^i_{a}$ as input and let $\mathtt{x}_a^\mu$ denote the output corresponding  to $\mathtt{u}^i_a$ as input.  The surface generated by the neural networks needs to be anchored to the boundary of the subregion provided. Let us select exactly $N_b$ discrete points on the boundary and denote them by $\mathtt{x}_a^{\text{bd}}$.  The differential equation for the bulk, denoted by $\mathcal{E}_{\text{bulk}}$ is given in equation \eqref{eq:bulkEOM} ). Let us denote the residual of the differential equation \eqref{eq:bulkEOM} at a point $x_a$ as $\mathtt{E}_{\text{bulk}}(x_a)$. With this notation, the loss function is written as
	\begin{equation}
		\label{eq:LHEE}
		\mathcal{L}_{\text{HEE}}=\frac{1}{N}\sum_{a=1}^N \mathtt{E}_{\text{bulk}}(x_a)^2+\frac{\lambda}{N_b}\sum_{a=1}^{N_b} (\mathtt{x}_a-\mathtt{x}^{\text{bd}}_a)^2~.
	\end{equation}
	The first sum imposes the condition for the differential equation to be satisfied at each of $N$ points and the second sum imposes the fact that the surface ends at the boundary specified in the problem. $\lambda$ denotes some real number which can be chosen appropriately to give more or less weightage to the boundary compared to the differential equation. 
    
    For our calculations in pure AdS, we chose the two hidden layers with 30 nodes each with $\tanh$ activation function in the intermediate layers and a $\text{softplus()}$ function at the end for the $z$ coordinate in order to obey $z \in [0,\infty]$. For the case of Blackhole with horizon radius $Z_h$ in asymptotically AdS, we replace $\text{softplus()}$ with $Z_h. \text{sigmoid}()$ for the $z$ coordinate since $z \in [0,Z_h]$. We found that keeping $\lambda$ in the range roughly from 100 to 1000 made the convergence better.
	
	The problem of finding EWCS is a constrained optimization problem where the minimal surface is constrained to lie on the RT surface connecting the subregions. The setup of the HEE goes through for most part except now we need to constrain the desired surface to end on the RT surface instead of just the boundary of a fixed subregion. Let us assume that the connected RT surface has been obtained using a neural network $\text{NN}_{\text{RT}}$ and the points are denoted by $x^{\RT}$. Next we define another neural network $\text{NN}_{\text{EWCS}_{bd}}$. It takes in $d-1$ inputs and gives $d$ outputs. The output of this network is fed in $\text{NN}_{\text{RT}}$. The composition of these two maps takes $d-1$ inputs and provides a point on the RT surface. This is meant to produce the boundary of EWCS constrained to live on the RT surface. To be concrete, consider the case of a 2-dimensional RT surface. In that case, the $\text{NN}_{\text{EWCS}_{bd}}$ takes in one input and upon composing it with $\text{NN}_{\text{RT}}$ provides points on the RT surface. Considering the domain of inputs to be a circle, this then gives us a 1-dimensional curve on the RT surface. For a general case, let us denote points on this curve as $\mathtt{x}^{\RT}_{a=1,\dots,N_{\RT}}$ for a choice of $N_{\RT}$ discrete points. To get the bulk of EWCS, we define a new network $\text{NN}_{\text{EWCS}}$ which takes $d$ inputs and gives $d+1$ outputs and it is used to define the bulk coordinates of ECWS. The networks $\text{NN}_{\text{EWCS}_{bd}}$ and $\text{NN}_{\text{EWCS}}$ are trained simultaneously. The loss function contains three terms, i.e., first term imposes differential equation in the bulk of EWCS, the second term imposes the orthogonality condition between the bulk of EWCS and the RT surface at the boundary of EWCS, and the third term imposes that the distance between predicted boundary of EWCS via $\text{NN}_{\text{EWCS}}$ and the boundary on the RT surface via composing $\text{NN}_{\text{EWCS}_{bd}}$ and $\text{NN}_{\text{EWCS}_{bd}}$ is vanishing. The second term of the loss function is described in equation \eqref{eq:boundaryEOM} and we denote the value of the differential equation a point $\mathtt{x}_a$ by $\mathtt{E}_{\text{boundary}}(\mathtt{x}_a)$. Finally, the loss function takes the following form
	\begin{equation}\label{eq:EWCSLoss}
		\mathcal{L}_{\text{EWCS}}=\frac{1}{N}\sum_{a=1}^N \mathtt{E}_{\text{bulk}}(x_a)^2+\frac{\lambda_1}{N_{\RT}}\sum_{a=1}^{N_{\RT}} \mathtt{E}_{\text{boundary}}(\mathtt{x}_a)^2+\frac{\lambda_2}{N_{\RT}}\sum_{a=1}^{N_{\RT}} (\mathtt{x}_a-\mathtt{x}^{\RT}_a)^2~.
	\end{equation}
    As in the case of equation \eqref{eq:LHEE}, we introduce $\lambda_1$ and $\lambda_2$ for each of the sums that impose boundary conditions.  In this case, we found that keeping $\lambda_{1}$ between $100$ to $1000$ and $\lambda_2 \simeq 10$ made the convergence better.
	
	A difference from the usual numerical methods is that the convergence to the answer depends on the learning rate, number of epochs and NN architecture. We have tried some variations of the architecture and have run the network with various epochs and learning rates. We choose the parameters to obtain numbers consistent with known results and then run the network for unknown results. We have also introduced $\lambda$, $\lambda_1$ and $\lambda_2$ in our loss functions. The choice of these coefficients affects how much weight a certain loss term is given and the convergence of the optimizer also depends on their choice. We experiment and choose them appropriately to achieve a desirable convergence rate.

    Finally, we note that the computation of area of the minimal surfaces after they are formed by the network requires some care. Since the metric in $\AdS_d$, behaves like $\sim 1/z^2$ as $z\to 0$, the computation of area depends very delicately on the choice of cutoff $z = \epsilon$. The network does not make a surface with the endpoints exactly on $z = \epsilon$. This is because the network tries to minimize the loss, but cannot make it zero, i.e., put the points exactly on the cutoff. In addition, the points mapped at the boundary don't have a constant $z$ coordinate. Therefore in order to find the area with a constant uniform cutoff, one has to find the input parameters, i.e., points in the domain that is fed into the neural network which correspond to a constant $z$ at the boundary. For this, we used the Newton-Raphson method. For different values of $z$ at the boundary, we can get a locus of domain points and one can then find the area of the surface for different cutoffs and study the cutoff dependence. In addition, to calculate the area with the cutoff, one needs to take a very fine grid, i.e., roughly of the order $20000\times 20000$ to get a good estimate of the actual area since the metric is divergent near $z=0$. This issue of cutoff, however, doesn't appear when we calculate the area of the EWCS since there is no cutoff dependence there.
    
	 \section{Differential equations for the extremal surface}
	\label{app:diffeqn}
	Let us do the computation in asymptotically $\AdS_{d+2}$ and denote the coordinates as $t$ and $X^{A=1,2,\dots d+2}$, with the metric denoted as $g_{AB}$. We restrict ourselves to a fixed time slice and parametrize the codimension-2 surface (denoted by $M$) by $\alpha^{i=1,2,\dots d+1}$. Let us denote the domain of $\alpha^i$ by $U$ and the induced metric $h$ on the hypersurface takes the form
	\begin{equation}
		h_{ij}=\frac{\partial X^A}{\partial \alpha^i}\frac{\partial X^B}{\partial \alpha^j}g_{AB}~.
	\end{equation}
	The area function takes the form 
	\begin{equation}
		S=\int_U \sqrt{h}~.
	\end{equation}

	The variation of this integral has a boundary and a bulk contribution
	\begin{equation}
		\delta S=\delta S_{\text{boundary}}+\delta S_{\text{bulk}}~,
	\end{equation}
	where
	\begin{equation}
		\delta S_{\text{bulk}}=-\int_U \left(	\frac{\partial}{\partial \alpha^i}\left(\sqrt{h} h^{ij}\frac{\partial X^A}{\partial \alpha^j }\right)g_{AC}+\frac{1}{2}\sqrt{h} h^{ij}\left(\frac{\partial g_{CB}}{\partial X^A}+\frac{\partial g_{AC}}{\partial X^B}-\frac{\partial g_{AB}}{\partial X^C}\right)\frac{\partial X^A}{\partial \alpha^i}\frac{\partial  X^B}{\partial \alpha^j}\right)\delta X^C
	\end{equation}
	and 
	\begin{equation}
		\delta S_{\text{boundary}}=\int_{\partial U}  \sqrt{h} n_i h^{ij}g_{AB}\frac{\partial X^B}{\partial \alpha^j}\delta \widetilde{X}^A~.
	\end{equation}
	Here $\delta\widetilde{X}^A$ denotes variation of the boundary endpoint and $n_i = d{\alpha}^{k_1} d{\alpha}^{k_2}...d{\alpha}^{k_{d-1}} \epsilon_{k_1k_2...k_{d-1}i}$ denotes the normal vector to the boundary of the $\alpha$ parametrization space. The bulk variation leads to the following equation
	\begin{equation}
		\label{eq:bulkEOM}\mathcal{E}_{\text{bulk}}=\frac{\partial}{\partial \alpha^i}\left(\sqrt{h} h^{ij}\frac{\partial X^A}{\partial \alpha^j }\right)+\sqrt{h} h^{ij}\Gamma^A_{BC} \frac{\partial X^B}{\partial \alpha^i}\frac{\partial  X^C}{\partial \alpha^j}=0
	\end{equation}
	For computing HEE, the boundary is a fixed shape and therefore $\delta \widetilde{X}^A=0$. For computing EWCS, we restrict the surface to lie on the RT surface. Let the parameters denoting this variation by $\beta^{a=1,2,\dots d}$.  This gives the following boundary conditions
	\begin{equation}
		\label{eq:boundaryEOM}\mathcal{E}_{\text{boundary}}=\sqrt{h} n_i h^{ij}g_{AB}\frac{\partial X^B}{\partial \alpha^j} \frac{\partial{\widetilde{X}}^A}{\partial \beta^a}=0
	\end{equation}
This condition essentially imposes orthogonality between the EWCS and the RT surface. It can be thought of as a generalization of the Neumann boundary condition.

    \bibliographystyle{JHEP}
	\bibliography{refversion2.bib}

@article{Ryu:2006bv,
	author = "Ryu, Shinsei and Takayanagi, Tadashi",
	title = "{Holographic derivation of entanglement entropy from AdS/CFT}",
	eprint = "hep-th/0603001",
	archivePrefix = "arXiv",
	reportNumber = "NSF-KITP-06-11",
	doi = "10.1103/PhysRevLett.96.181602",
	journal = "Phys. Rev. Lett.",
	volume = "96",
	pages = "181602",
	year = "2006"
}

@article{Ryu:2006ef,
	author = "Ryu, Shinsei and Takayanagi, Tadashi",
	title = "{Aspects of Holographic Entanglement Entropy}",
	eprint = "hep-th/0605073",
	archivePrefix = "arXiv",
	reportNumber = "NSF-KITP-06-31, KUNS-2021",
	doi = "10.1088/1126-6708/2006/08/045",
	journal = "JHEP",
	volume = "08",
	pages = "045",
	year = "2006"
}

@article{Astaneh:2014uba,
	author = "Astaneh, Amin Faraji and Gibbons, Gary and Solodukhin, Sergey N.",
	title = "{What surface maximizes entanglement entropy?}",
	eprint = "1407.4719",
	archivePrefix = "arXiv",
	primaryClass = "hep-th",
	doi = "10.1103/PhysRevD.90.085021",
	journal = "Phys. Rev. D",
	volume = "90",
	number = "8",
	pages = "085021",
	year = "2014"
}

@article{Allais_2015,
	title={Some results on the shape dependence of entanglement and Rényi entropies},
	volume={91},
	ISSN={1550-2368},
	url={http://dx.doi.org/10.1103/PhysRevD.91.046002},
	DOI={10.1103/physrevd.91.046002},
	number={4},
	journal={Physical Review D},
	publisher={American Physical Society (APS)},
	author={Allais, Andrea and Mezei, Márk},
	year={2015},
	month=feb }

@article{Fonda:2015nma,
	author = "Fonda, Piermarco and Seminara, Domenico and Tonni, Erik",
	title = "{On shape dependence of holographic entanglement entropy in AdS$_{4}$/CFT$_{3}$}",
	eprint = "1510.03664",
	archivePrefix = "arXiv",
	primaryClass = "hep-th",
	doi = "10.1007/JHEP12(2015)037",
	journal = "JHEP",
	volume = "12",
	pages = "037",
	year = "2015"
}

@article{Klebanov:2012yf,
	author = "Klebanov, Igor R. and Nishioka, Tatsuma and Pufu, Silviu S. and Safdi, Benjamin R.",
	title = "{On Shape Dependence and RG Flow of Entanglement Entropy}",
	eprint = "1204.4160",
	archivePrefix = "arXiv",
	primaryClass = "hep-th",
	reportNumber = "PUPT-2411, MIT-CTP-4357",
	doi = "10.1007/JHEP07(2012)001",
	journal = "JHEP",
	volume = "07",
	pages = "001",
	year = "2012"
}

@article{Carmi:2015dla,
	author = "Carmi, Dean",
	title = "{On the Shape Dependence of Entanglement Entropy}",
	eprint = "1506.07528",
	archivePrefix = "arXiv",
	primaryClass = "hep-th",
	doi = "10.1007/JHEP12(2015)043",
	journal = "JHEP",
	volume = "12",
	pages = "043",
	year = "2015"
}

@article{Lewkowycz:2018sgn,
	author = "Lewkowycz, Aitor and Parrikar, Onkar",
	title = "{The holographic shape of entanglement and Einstein\textquoteright{}s equations}",
	eprint = "1802.10103",
	archivePrefix = "arXiv",
	primaryClass = "hep-th",
	doi = "10.1007/JHEP05(2018)147",
	journal = "JHEP",
	volume = "05",
	pages = "147",
	year = "2018"
}

@article{Bueno:2015xda,
	author = "Bueno, Pablo and Myers, Robert C.",
	title = "{Corner contributions to holographic entanglement entropy}",
	eprint = "1505.07842",
	archivePrefix = "arXiv",
	primaryClass = "hep-th",
	doi = "10.1007/JHEP08(2015)068",
	journal = "JHEP",
	volume = "08",
	pages = "068",
	year = "2015"
}

@article{Umemoto_2018,
	title={Entanglement of purification through holographic duality},
	volume={14},
	ISSN={1745-2481},
	url={http://dx.doi.org/10.1038/s41567-018-0075-2},
	DOI={10.1038/s41567-018-0075-2},
	number={6},
	journal={Nature Physics},
	publisher={Springer Science and Business Media LLC},
	author={Umemoto, Koji and Takayanagi, Tadashi},
	year={2018},
	month=mar, pages={573–577} }

@article{Jokela:2019ebz,
	author = {Jokela, Niko and P\"onni, Arttu},
	title = "{Notes on entanglement wedge cross sections}",
	eprint = "1904.09582",
	archivePrefix = "arXiv",
	primaryClass = "hep-th",
	reportNumber = "HIP-2019-10/TH",
	doi = "10.1007/JHEP07(2019)087",
	journal = "JHEP",
	volume = "07",
	pages = "087",
	year = "2019"
}

@article{BabaeiVelni:2019pkw,
	author = "Babaei Velni, Komeil and Mohammadi Mozaffar, M. Reza and Vahidinia, M. H.",
	title = "{Some Aspects of Entanglement Wedge Cross-Section}",
	eprint = "1903.08490",
	archivePrefix = "arXiv",
	primaryClass = "hep-th",
	reportNumber = "IPM/P-2019/007",
	doi = "10.1007/JHEP05(2019)200",
	journal = "JHEP",
	volume = "05",
	pages = "200",
	year = "2019"
}

@article{em/1048709050,
	author = {Kenneth A. Brakke},
	title = {{The surface evolver}},
	volume = {1},
	journal = {Experimental Mathematics},
	number = {2},
	publisher = {A K Peters, Ltd.},
	pages = {141 -- 165},
	year = {1992},
}

@article{Cavini:2019wyb,
	author = "Cavini, Giacomo and Seminara, Domenico and Sisti, Jacopo and Tonni, Erik",
	title = "{On shape dependence of holographic entanglement entropy in AdS$_{4}$/CFT$_{3}$ with Lifshitz scaling and hyperscaling violation}",
	eprint = "1907.10030",
	archivePrefix = "arXiv",
	primaryClass = "hep-th",
	doi = "10.1007/JHEP02(2020)172",
	journal = "JHEP",
	volume = "02",
	pages = "172",
	year = "2020"
}

@article{Seminara:2017hhh,
	author = "Seminara, Domenico and Sisti, Jacopo and Tonni, Erik",
	title = "{Corner contributions to holographic entanglement entropy in AdS$_{4}$/BCFT$_{3}$}",
	eprint = "1708.05080",
	archivePrefix = "arXiv",
	primaryClass = "hep-th",
	doi = "10.1007/JHEP11(2017)076",
	journal = "JHEP",
	volume = "11",
	pages = "076",
	year = "2017"
}

@article{Fonda:2014cca,
	author = "Fonda, Piermarco and Giomi, Luca and Salvio, Alberto and Tonni, Erik",
	title = "{On shape dependence of holographic mutual information in AdS$_{4}$}",
	eprint = "1411.3608",
	archivePrefix = "arXiv",
	primaryClass = "hep-th",
	reportNumber = "IFT-UAM-CSIC-14-118",
	doi = "10.1007/JHEP02(2015)005",
	journal = "JHEP",
	volume = "02",
	pages = "005",
	year = "2015"
}

@inproceedings{Kingma:2014vow,
	author = "Kingma, Diederik P. and Ba, Jimmy",
	title = "{Adam: A Method for Stochastic Optimization}",
	eprint = "1412.6980",
	archivePrefix = "arXiv",
	primaryClass = "cs.LG",
	month = "12",
	year = "2014"
}

@article{Cybenko:1989iql,
	author = "Cybenko, G.",
	title = "{Approximation by superpositions of a sigmoidal function}",
	doi = "10.1007/BF02551274",
	journal = "Math. Control Signals Syst.",
	volume = "2",
	number = "4",
	pages = "303--314",
	year = "1989"
}

@article{Halverson:2024hax,
	author = "Halverson, Jim",
	title = "{TASI Lectures on Physics for Machine Learning}",
	eprint = "2408.00082",
	archivePrefix = "arXiv",
	primaryClass = "hep-th",
	month = "7",
	year = "2024"
}

@article{Terhal:2002riz,
	author = "Terhal, Barbara M. and Horodecki, Michal and Leung, Debbie W. and DiVincenzo, David P.",
	title = "{The entanglement of purification}",
	eprint = "quant-ph/0202044",
	archivePrefix = "arXiv",
	doi = "10.1063/1.1498001",
	journal = "J. Math. Phys.",
	volume = "43",
	number = "9",
	pages = "4286--4298",
	year = "2002"
}

@mastersthesis{Chopp:1991uf,
	author = "Chopp, David L.",
	title = "{Computing minimal surfaces via level set curvature flow}",
	reportNumber = "LBL-30685",
	type = "Other thesis",
	month = "5",
	year = "1991"
}

@article{Mezei:2014zla,
	author = "Mezei, M{\'a}rk",
	title = "{Entanglement entropy across a deformed sphere}",
	eprint = "1411.7011",
	archivePrefix = "arXiv",
	primaryClass = "hep-th",
	doi = "10.1103/PhysRevD.91.045038",
	journal = "Phys. Rev. D",
	volume = "91",
	number = "4",
	pages = "045038",
	year = "2015"
}

@article{Nishioka:2018khk,
	author = "Nishioka, Tatsuma",
	title = "{Entanglement entropy: holography and renormalization group}",
	eprint = "1801.10352",
	archivePrefix = "arXiv",
	primaryClass = "hep-th",
	reportNumber = "UT-18-02",
	doi = "10.1103/RevModPhys.90.035007",
	journal = "Rev. Mod. Phys.",
	volume = "90",
	number = "3",
	pages = "035007",
	year = "2018"
}

@article{Headrick:2019eth,
	author = "Headrick, Matthew",
	title = "{Lectures on entanglement entropy in field theory and holography}",
	eprint = "1907.08126",
	archivePrefix = "arXiv",
	primaryClass = "hep-th",
	reportNumber = "BRX-TH-6653",
	month = "7",
	year = "2019"
}

@inproceedings{VanRaamsdonk:2016exw,
	author = "Van Raamsdonk, Mark",
	title = "{Lectures on gravity and entanglement.}",
	booktitle = "{Theoretical Advanced Study Institute in Elementary Particle Physics}: {New Frontiers in Fields and Strings}",
	eprint = "1609.00026",
	archivePrefix = "arXiv",
	primaryClass = "hep-th",
	doi = "10.1142/9789813149441_0005",
	pages = "297--351",
	year = "2017"
}

@article{Casini:2009sr,
	author = "Casini, H. and Huerta, M.",
	title = "{Entanglement entropy in free quantum field theory}",
	eprint = "0905.2562",
	archivePrefix = "arXiv",
	primaryClass = "hep-th",
	doi = "10.1088/1751-8113/42/50/504007",
	journal = "J. Phys. A",
	volume = "42",
	pages = "504007",
	year = "2009"
}

@article{Calabrese:2009qy,
	author = "Calabrese, Pasquale and Cardy, John",
	title = "{Entanglement entropy and conformal field theory}",
	eprint = "0905.4013",
	archivePrefix = "arXiv",
	primaryClass = "cond-mat.stat-mech",
	doi = "10.1088/1751-8113/42/50/504005",
	journal = "J. Phys. A",
	volume = "42",
	pages = "504005",
	year = "2009"
}

@article{Nishioka:2009un,
	author = "Nishioka, Tatsuma and Ryu, Shinsei and Takayanagi, Tadashi",
	title = "{Holographic Entanglement Entropy: An Overview}",
	eprint = "0905.0932",
	archivePrefix = "arXiv",
	primaryClass = "hep-th",
	reportNumber = "KUNS-2207, IPMU09-0056",
	doi = "10.1088/1751-8113/42/50/504008",
	journal = "J. Phys. A",
	volume = "42",
	pages = "504008",
	year = "2009"
}

@article{Witten:2018zxz,
	author = "Witten, Edward",
	title = "{APS Medal for Exceptional Achievement in Research: Invited article on entanglement properties of quantum field theory}",
	eprint = "1803.04993",
	archivePrefix = "arXiv",
	primaryClass = "hep-th",
	doi = "10.1103/RevModPhys.90.045003",
	journal = "Rev. Mod. Phys.",
	volume = "90",
	number = "4",
	pages = "045003",
	year = "2018"
}

@article{Ahn:2024jkk,
	author = "Ahn, Byoungjoon and Jeong, Hyun-Sik and Kim, Keun-Young and Yun, Kwan",
	title = "{Holographic reconstruction of black hole spacetime: machine learning and entanglement entropy}",
	eprint = "2406.07395",
	archivePrefix = "arXiv",
	primaryClass = "hep-th",
	reportNumber = "IFT-UAM/CSIC-24-88",
	doi = "10.1007/JHEP01(2025)025",
	journal = "JHEP",
	volume = "01",
	pages = "025",
	year = "2025"
}

@article{Takayanagi:2025ula,
	author = "Takayanagi, Tadashi",
	title = "{Essay: Emergent Holographic Spacetime from Quantum Information}",
	eprint = "2506.06595",
	archivePrefix = "arXiv",
	primaryClass = "hep-th",
	reportNumber = "YITP-25-58",
	doi = "10.1103/pg4r-fy8n",
	journal = "Phys. Rev. Lett.",
	volume = "134",
	number = "24",
	pages = "240001",
	year = "2025"
}

@article{Carleo:2019ptp,
	author = "Carleo, Giuseppe and Cirac, Ignacio and Cranmer, Kyle and Daudet, Laurent and Schuld, Maria and Tishby, Naftali and Vogt-Maranto, Leslie and Zdeborov{\'a}, Lenka",
	title = "{Machine learning and the physical sciences}",
	eprint = "1903.10563",
	archivePrefix = "arXiv",
	primaryClass = "physics.comp-ph",
	doi = "10.1103/RevModPhys.91.045002",
	journal = "Rev. Mod. Phys.",
	volume = "91",
	number = "4",
	pages = "045002",
	year = "2019"
}

@article{Mehta:2018dln,
	author = "Mehta, Pankaj and Bukov, Marin and Wang, Ching-Hao and Day, Alexandre G. R. and Richardson, Clint and Fisher, Charles K. and Schwab, David J.",
	title = "{A high-bias, low-variance introduction to Machine Learning for physicists}",
	eprint = "1803.08823",
	archivePrefix = "arXiv",
	primaryClass = "physics.comp-ph",
	doi = "10.1016/j.physrep.2019.03.001",
	journal = "Phys. Rept.",
	volume = "810",
	pages = "1--124",
	year = "2019"
}

@article{karniadakis2021physics,
	title={Physics-informed machine learning},
	author={Karniadakis, George Em and Kevrekidis, Ioannis G and Lu, Lu and Perdikaris, Paris and Wang, Sifan and Yang, Liu},
	journal={Nature Reviews Physics},
	volume={3},
	number={6},
	pages={422--440},
	year={2021},
	publisher={Nature Publishing Group UK London}
}

@article{cuomo2022scientific,
	title={Scientific machine learning through physics--informed neural networks: Where we are and what’s next},
	author={Cuomo, Salvatore and Di Cola, Vincenzo Schiano and Giampaolo, Fabio and Rozza, Gianluigi and Raissi, Maziar and Piccialli, Francesco},
	journal={Journal of Scientific Computing},
	volume={92},
	number={3},
	pages={88},
	year={2022},
	publisher={Springer}
}

@article{de2024numerical,
	title={Numerical analysis of physics-informed neural networks and related models in physics-informed machine learning},
	author={De Ryck, Tim and Mishra, Siddhartha},
	journal={Acta Numerica},
	volume={33},
	pages={633--713},
	year={2024},
	publisher={Cambridge University Press}
}

@article{Schmidhuber:2014bpo,
	author = {Schmidhuber, J{\"u}rgen},
	title = "{Deep learning in neural networks: An overview}",
	doi = "10.1016/j.neunet.2014.09.003",
	journal = "Neural Networks",
	volume = "61",
	pages = "85--117",
	year = "2015"
}

@article{LeCun:2015pmr,
	author = "LeCun, Yann and Bengio, Yoshua and Hinton, Geoffrey",
	title = "{Deep learning}",
	doi = "10.1038/nature14539",
	journal = "Nature",
	volume = "521",
	pages = "436--444",
	year = "2015"
}

@book{bishop2006pattern,
	title={Pattern recognition and machine learning},
	author={Bishop, Christopher M and Nasrabadi, Nasser M},
	volume={4},
	number={4},
	year={2006},
	publisher={Springer}
}

@book{goodfellow2016deep,
	title={Deep learning},
	author={Goodfellow, Ian and Bengio, Yoshua and Courville, Aaron and Bengio, Yoshua},
	volume={1},
	number={2},
	year={2016},
	publisher={MIT press Cambridge}
}

@article{raissi2017physics,
	title={Physics informed deep learning (part i): Data-driven solutions of nonlinear partial differential equations},
	author={Raissi, Maziar and Perdikaris, Paris and Karniadakis, George Em},
	journal={arXiv preprint arXiv:1711.10561},
	year={2017}
}

@article{raissi2019physics,
	title={Physics-informed neural networks: A deep learning framework for solving forward and inverse problems involving nonlinear partial differential equations},
	author={Raissi, Maziar and Perdikaris, Paris and Karniadakis, George E},
	journal={Journal of Computational physics},
	volume={378},
	pages={686--707},
	year={2019},
	publisher={Elsevier}
}

@article{Nguyen:2017yqw,
	author = "Nguyen, Phuc and Devakul, Trithep and Halbasch, Matthew G. and Zaletel, Michael P. and Swingle, Brian",
	title = "{Entanglement of purification: from spin chains to holography}",
	eprint = "1709.07424",
	archivePrefix = "arXiv",
	primaryClass = "hep-th",
	doi = "10.1007/JHEP01(2018)098",
	journal = "JHEP",
	volume = "01",
	pages = "098",
	year = "2018"
}

@article{Hashimoto:2025upw,
	author = "Hashimoto, Koji and Kyo, Koichi and Murata, Masaki and Ogiwara, Gakuto and Tanahashi, Norihiro",
	title = "{Gluon scattering amplitudes with instantons and minimal surfaces with topology change}",
	eprint = "2509.10865",
	archivePrefix = "arXiv",
	primaryClass = "hep-th",
	reportNumber = "KUNS-3071",
	month = "9",
	year = "2025"
}

@article{Liu:1989esw,
	author = "Liu, Dong C. and Nocedal, Jorge",
	title = "{On the limited memory BFGS method for large scale optimization}",
	doi = "10.1007/BF01589116",
	journal = "Math. Programming",
	volume = "45",
	number = "1",
	pages = "503--528",
	year = "1989"
}

@article{Paszke:2019xhz,
	author = "Paszke, Adam and others",
	title = "{PyTorch: An Imperative Style, High-Performance Deep Learning Library}",
	eprint = "1912.01703",
	archivePrefix = "arXiv",
	primaryClass = "cs.LG",
	month = "12",
	year = "2019"
}

@article{Lam:2021ugb,
	author = "Lam, Jonathan and You, Yi-Zhuang",
	title = "{Machine learning statistical gravity from multi-region entanglement entropy}",
	eprint = "2110.01115",
	archivePrefix = "arXiv",
	primaryClass = "hep-th",
	doi = "10.1103/PhysRevResearch.3.043199",
	journal = "Phys. Rev. Res.",
	volume = "3",
	number = "4",
	pages = "043199",
	year = "2021"
}

@article{Park:2022fqy,
	author = "Park, Chanyong and Hwang, Chi-Ok and Cho, Kyungchan and Kim, Se-Jin",
	title = "{Dual geometry of entanglement entropy via deep learning}",
	eprint = "2205.04445",
	archivePrefix = "arXiv",
	primaryClass = "hep-th",
	doi = "10.1103/PhysRevD.106.106017",
	journal = "Phys. Rev. D",
	volume = "106",
	number = "10",
	pages = "106017",
	year = "2022"
}

@article{Bulgarelli:2024yrz,
	author = {Bulgarelli, Andrea and Cellini, Elia and Jansen, Karl and K{\"u}hn, Stefan and Nada, Alessandro and Nakajima, Shinichi and Nicoli, Kim A. and Panero, Marco},
	title = "{Flow-Based Sampling for Entanglement Entropy and the Machine Learning of Defects}",
	eprint = "2410.14466",
	archivePrefix = "arXiv",
	primaryClass = "quant-ph",
	doi = "10.1103/PhysRevLett.134.151601",
	journal = "Phys. Rev. Lett.",
	volume = "134",
	number = "15",
	pages = "151601",
	year = "2025"
}

@article{Bueno:2021fxb,
	author = "Bueno, Pablo and Casini, Horacio and Andino, Oscar Lasso and Moreno, Javier",
	title = "{Disks globally maximize the entanglement entropy in 2 + 1 dimensions}",
	eprint = "2107.12394",
	archivePrefix = "arXiv",
	primaryClass = "hep-th",
	doi = "10.1007/JHEP10(2021)179",
	journal = "JHEP",
	volume = "10",
	pages = "179",
	year = "2021"
}

@article{Faulkner:2015csl,
	author = "Faulkner, Thomas and Leigh, Robert G. and Parrikar, Onkar",
	title = "{Shape Dependence of Entanglement Entropy in Conformal Field Theories}",
	eprint = "1511.05179",
	archivePrefix = "arXiv",
	primaryClass = "hep-th",
	doi = "10.1007/JHEP04(2016)088",
	journal = "JHEP",
	volume = "04",
	pages = "088",
	year = "2016"
}

@article{Ahn:2024gjf,
    author = "Ahn, Byoungjoon and Jeong, Hyun-Sik and Kim, Keun-Young and Yun, Kwan",
    title = "{Deep learning bulk spacetime from boundary optical conductivity}",
    eprint = "2401.00939",
    archivePrefix = "arXiv",
    primaryClass = "hep-th",
    reportNumber = "IFT-UAM/CSIC-24-24",
    doi = "10.1007/JHEP03(2024)141",
    journal = "JHEP",
    volume = "03",
    pages = "141",
    year = "2024"
}

@article{Ahn:2025tjp,
    author = "Ahn, Byoungjoon and Jeong, Hyun-Sik and Ji, Chang-Woo and Kim, Keun-Young and Yun, Kwan",
    title = "{Deep learning-based holography for T-linear resistivity}",
    eprint = "2502.10245",
    archivePrefix = "arXiv",
    primaryClass = "hep-th",
    reportNumber = "IFT-UAM/CSIC-25-2",
    month = "2",
    year = "2025"
}

@article{Jeong:2019xdr,
    author = "Jeong, Hyun-Sik and Kim, Keun-Young and Nishida, Mitsuhiro",
    title = "{Reflected Entropy and Entanglement Wedge Cross Section with the First Order Correction}",
    eprint = "1909.02806",
    archivePrefix = "arXiv",
    primaryClass = "hep-th",
    doi = "10.1007/JHEP12(2019)170",
    journal = "JHEP",
    volume = "12",
    pages = "170",
    year = "2019"
}

@article{Hashimoto:2025zmiupd,
    author = "Hashimoto, Koji and Kyo, Koichi and Murata, Masaki and Ogiwara, Gakuto and Tanahashi, Norihiro",
    title = "{Physics-informed neural network solves minimal surfaces in curved spacetime}",
    eprint = "2509.10866",
    archivePrefix = "arXiv",
    primaryClass = "hep-th",
    reportNumber = "KUNS-3072",
    doi = "10.1088/2632-2153/ae3050",
    journal = "Mach. Learn. Sci. Tech.",
    volume = "7",
    number = "1",
    pages = "015013",
    year = "2026"
}
\end{document}